\documentclass[12pt,draftclsnofoot, onecolumn]{IEEEtran}

\ifCLASSINFOpdf
\else
\fi
\usepackage{amsmath,graphicx}
\usepackage{subfigure}
\usepackage{lineno}
\usepackage{cite}
\usepackage{multirow}
\usepackage{stfloats}
\usepackage{amsfonts}
\usepackage{amssymb}
\usepackage{booktabs} 
\usepackage{makecell}
\usepackage{diagbox}
\usepackage{slashbox}
\usepackage{color}
\usepackage[ruled,linesnumbered,boxed]{algorithm2e}
\begin{document}
	%
	\title{Time-Varying Channel Prediction for RIS-Assisted MU-MISO Networks via Deep Learning }
	%
	%
	%
	
	
	\author{
		\IEEEauthorblockN{
			Wangyang Xu, Jiancheng An,  Yongjun Xu, \IEEEmembership{Senior Member, IEEE}, Chongwen Huang, Lu Gan, and Chau Yuen, \IEEEmembership{Fellow, IEEE}
			\vspace{-3em}}
		\thanks{W. Xu, J. An, and L. Gan are with the School of Information and Communication Engineering, University of Electronic Science and Technology of China, Chengdu, Sichuan, 611731, China (e-mail: wangyangxu@std.uestc.edu.cn; jiancheng\_an@163.com; ganlu@uestc.edu.cn).}
		\thanks {Y. Xu is with the School of Communication and Information Engineering, Chongqing Key Laboratory of Mobile Communications Technology, Chongqing University of Posts and Telecommunications, Chongqing 400065, China (e-mail: xuyj@cqupt.edu.cn).}
		\thanks{C. Huang is with the College of Information Science and Electronic Engineering, Zhejiang University, Hangzhou 310027, China, and with International Joint Innovation Center, Zhejiang University, Haining 314400, China, and also with Zhejiang Provincial Key Laboratory of Info. Proc., Commun. \& Netw. (IPCAN), Hangzhou 310027, China. (e-mail: chongwenhuang@zju.edu.cn).}
		\thanks{C. Yuen is with Engineering Product Development (EPD) Pillar, Singapore University of Technology and Design, Singapore 487372, Singapore (e-mail: yuenchau@sutd.edu.sg).}
	}
	\maketitle
	
	\begin{abstract}
		To mitigate the effects of shadow fading and obstacle blocking, reconfigurable intelligent surface (RIS) has become a promising technology to improve the signal transmission quality of wireless communications by controlling the reconfigurable passive elements with less hardware cost and lower power consumption. However, accurate, low-latency and low-pilot-overhead channel state information (CSI) acquisition remains a considerable challenge in RIS-assisted systems due to the large number of RIS passive elements. In this paper, we propose a three-stage joint channel decomposition and prediction framework to require CSI. The proposed framework exploits the two-timescale property that the base station (BS)-RIS channel is quasi-static and the RIS-user equipment (UE) channel is fast time-varying. Specifically, in the first stage, we use the full-duplex technique to estimate the channel between a BS's specific antenna and the RIS, addressing the critical scaling ambiguity problem in the channel decomposition. We then design a novel deep neural network, namely, the sparse-connected long short-term memory (SCLSTM), and propose a SCLSTM-based algorithm in the second and third stages, respectively. The algorithm can simultaneously decompose the BS-RIS channel and RIS-UE channel from the cascaded channel and capture the temporal relationship of the RIS-UE channel for prediction. Simulation results show that our proposed framework has lower pilot overhead than the traditional channel estimation algorithms, and the proposed SCLSTM-based algorithm can also achieve more accurate CSI acquisition robustly and effectively.

	\end{abstract}
	
	\begin{IEEEkeywords}
		Reconfigurable intelligent surface, joint channel decomposition and prediction, deep learning, pilot overhead.
	\end{IEEEkeywords}

	%
	\IEEEpeerreviewmaketitle

	\section{Introduction}
	%
	%
	%
	%
	The demand for future B5G and 6G wireless communications is to make wireless communication systems more high-speed, low-power, energy-efficient, and low-latency with the explosion of device connectivity and information exchange \cite{B5G,6G2}. Although existing technologies in wireless communications, such as massive multiple-input multiple-output, can improve spectrum efficiency and system throughput, the transmission between the transmitter and the receiver might be affected by the lousy transmission environments \cite{larsson2014massive}. Another technology that can overcome the influence of the awful environment is the relay communication (i.e., cooperative communication), which requires more hardware consumption and complex signal processing \cite{liangrelay,neikato2}.

	Recently, reconfigurable intelligent surface (RIS) is known as a potential technology that dramatically improves system throughput, spectrum efficiency, and energy efficiency of wireless communications \cite{huangTWC,liangyingchang1,wu2019beamforming,Hoang1,WCT,huang2020holographic,outage,CST,zhiqin}. An RIS is equipped with a large number of hardware-efficient and nearly passive reflecting elements. Thanks to the development of programmable materials, the phase shifts of these elements at the RIS can be adjusted in real time to achieve intelligent manipulation of the communication environment and thus improve communication quality \cite{gaofeifei1}. 
	
	There have been many works on the performance optimization of RIS-assisted wireless communications. For example, the authors in \cite{huangTWC} and \cite{energyefficient} analyzed  the energy efficiency potential of RIS in the uplink multi-user multiple-input multiple-output (MU-MISO) communications. The capacity improvement can be achieved through a well-designed precoding matrix at the base station (BS) and the passive reflection coefficient vector at the RIS under different constraints. Specifically, the authors in \cite{wujointcontinuous,guo2020weighted,10MISOCEBO,an1,an2} solved the joint beamforming problem under the assumption of continuous phase shifts, while the practical discrete phase shifts were considered in \cite{13xu,14xu,15xu,di2020hybrid}. Besides, the combinations of RIS-assisted systems with other technologies such as non-orthogonal multiple access, millimeter wave (mmWave), and Terahertz (THz) have been investigated in \cite{NOMA,mmwavethz}. The advantages of RIS-assisted systems are achieved by a reasonable design for the RIS, which requires obtaining the channel state information (CSI) of the system in advance. 
	\subsection{Prior Works}
	Most of the existing works on the CSI acquisition of RIS-assisted systems have focused on pilot-assisted channel estimation to estimate the cascaded channel of the BS-RIS-user equipment (UE). For example, both \cite{cascaded1} and \cite{cascaded2} estimated the cascaded channel by switching on and off the reflective element of the RIS during different time slots. The difference is that the former used a minimum mean square error criterion, while the latter combined sparse matrix decomposition and matrix completion methods. However, this scheme suffers from the degradation of the signal-to-noise ratio (SNR) since only one element of the RIS reflects the pilots during a time slot. In \cite{MVU}, the authors proposed an optimal channel estimation scheme based on a minimum variance unbiased estimator by designing a series of reflection coefficient vectors. In \cite{BSRIS}, the authors assumed that the BS-RIS channels between different users were constant. Then, they estimated the cascaded channel of the first user as a reference channel for estimating the cascaded channels of other users to reduce the pilot overhead. Nevertheless, the three-phase method of \cite{BSRIS} suffers from severe error propagation, especially at the low SNR region. In \cite{CStradition}, the authors further reduce the pilot overhead upon leveraging the low-rank property of the cascaded channel. In \cite{CEhuang}, the authors estimated the separated BS-RIS and RIS-UE channels from the cascaded channel model based on a PARAllel FACtor (PARAFAC) decomposition. In addition, deep learning methods have also been applied in channel estimation to reduce the protocol overhead \cite{CEDL1,CEDL2,CEDL3,CEDL4}. However, most of their principles are based on channel sparsity and compressive sensing, which require a certain number of active elements to be configured at the RIS.
	
	Since the vast number of elements at the RIS cannot receive and transmit signals, the pilot overhead is unbearable in the pilot-based channel estimation methods. Especially in some  ultra-high bands situation such as mmWave and THz, where the coherence time of the channel is significantly reduced. In addition, the pilot-based channel estimation methods are no longer sufficient for the requirements of real-time CSI acquisition in high-mobility wireless communications because the channel coherence time is not enough for pilot transmission.
	
	Channel prediction is a technology that uses previous CSI to forecast future CSI. Unlike the pilot-based channel estimation algorithms, it can monitor the time-varying characteristic of the channel without any pilots to effectively solve the time-varying channel acquisition problem. Traditional channel prediction is usually based on the autoregressive predictive model \cite{CP1,CP2}, or the deterministic parameter-based model \cite{CP3}. However, channel prediction algorithms based on these models are not applicable for the  non-stationary and fast-varying environments. Although adaptive filtering techniques (e.g., Kalman filtering) can predict time-varying fast fading channels with the second-order statistics of the time-varying channel, and the cost is relatively high \cite{CP4}.
	\subsection{Contributions}
	To acquire the CSI of the time-varying channels in a RIS-assisted MU-MISO system, we propose a three-stage joint channel decomposition and prediction framework based on the two-timescale property and the channel prediction in this paper. Besides, we design a novel neural network (NN) structure, called sparse-connected long short-term memory (SCLSTM), and develop a SCLSTM-based algorithm to achieve channel decomposition and prediction. The contributions are summarized as follows:
	\begin{itemize}
		\item We leverage the two-timescale channel property, where the BS-RIS channel is quasi-static and the RIS-UE channel is time-varying. Then we propose a three-stage joint channel decomposition and prediction framework based on the two-timescale property to obtain the CSI in real time in a fast time-varying environment. In the first stage, we use a dual-link pilot transmission scheme to estimate the channel between the specific BS's antenna and the RIS to correct the scaling ambiguity in the subsequent channel decomposition. In the second stage, the initial channels are estimated for the channel prediction of the third stage, and the channel decomposition and prediction are completed in the third stage. Unlike pilot-based channel estimation, the proposed framework does not send pilots during the channel prediction.  
		\item We design a NN framework SCLSTM based on long short-term memory (LSTM), which can simultaneously perform channel decomposition and prediction. In the channel decomposition part, we construct two sparse-connected layers according to the mapping relationship between the input and the output, which significantly reduces the computational complexity of the NN. In the prediction part, the LSTM structure is used to capture the temporal relationship of the time-varying RIS-UE channel.
		\item We develop a SCLSTM-based algorithm for the proposed three-stage framework. Moreover, through the pilot-overhead analysis, computational-complexity analysis, and simulation results, we prove the effectiveness and robustness of the proposed framework and algorithm. The simulation results indicate that our proposed algorithm can be directly applied when the channel is fast time-varying and the channel coherence time is less than the period needed for pilots of the traditional pilot-based channel estimation algorithms. 
	\end{itemize}
	\subsection{Organization and Notations}
	The rest of this paper is organized as follows. Section II introduces the system and signal models of the RIS-assisted MU-MISO communication system. In Sections III, we propose the joint channel decomposition and prediction framework based on the two-timescale property. Section IV presents the SCLSTM-based algorithm for the proposed framework. Furthermore, the pilot overhead and the computational complexity are analyzed in Section V. Section VI presents the simulation results of our proposed algorithm. Finally, we conclude the paper in Section VII.
	
	\emph{Notations:} Scalars are denoted by italic letters. Vectors and matrices are denoted by bold-face lower-case and upper-case letters, respectively. The superscripts ${\left(  \cdot  \right)^T}$ and ${\left(  \cdot  \right)^H}$ denote the operations of transpose and Hermitian transpose. $\mathbb{E}{\left[  \cdot  \right]}$ denotes the statistical expectation. $\left|  \cdot  \right|$ denotes the absolute value of a real number. $\left\|  \cdot  \right\|$ denotes the 2-Norm of a vector or a matrix. $\operatorname{Re} \left\{  x  \right\}$ and $\operatorname{Im}  \left\{  x  \right\}$ denote the real and imaginary parts of $x$, respectively. ${\rm{diag}}\left(  \cdot  \right)$ denotes the diagonal operation. $\lceil  x  \rceil $ denotes the smallest integer that is greater than or equal to $x$. $ \odot $ is the Hadamard product.

	\section{System and Channel Models}
	\subsection{System Model}
\begin{figure}[tbp]
		\centering {
			\begin{tabular}{ccc}
				\includegraphics[width=0.7\textwidth]{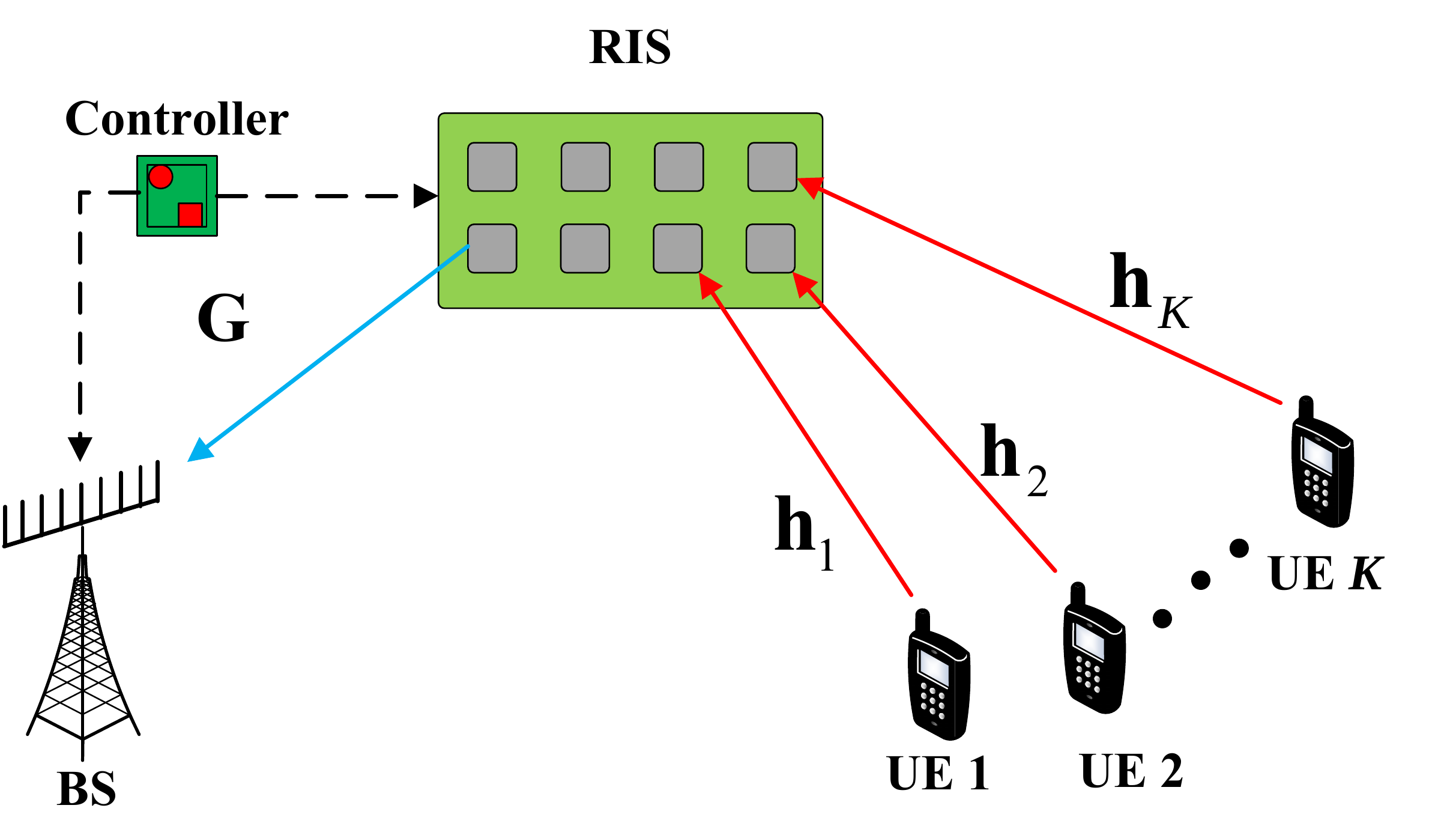}\\
			\end{tabular}
		}
		\caption{A RIS-assisted MU-MISO system.}
		\vspace{-1\baselineskip}
		\label{fig1}
	\end{figure}

	In this paper, we consider an uplink RIS-assisted mmWave MU-MISO system, as illustrated in Fig. \ref{fig1}. The system consists of a BS with $M$ antennas, $K$ single-antenna UEs, and one RIS equipped with $N$ passive reflecting elements in a rectangular arrangement. Defining ${\cal K} = \{ {1,2, \cdots K} \}$ as the UE set and $ {\forall k \in {\cal K}} $ in the following. Note that we ignore the direct links between the BS and UEs due to the high attenuation characteristics of mmWave and the harsh propagation environment. 
	
	In the uplink transmission, the received signal at the BS in the $t$-th time slot can be expressed as  
	\begin{align}   \label{xu1}
	\left.{\mathbf{y}}_t = \sum\limits_{k = 1}^K {{{\mathbf{G}}}{\boldsymbol{\Theta}}_t {{{{\mathbf{h}}_{k}}}}{{{{x}}_{k,t}}} + {{\mathbf{n}}_t}},\right.
	\end{align}
	where ${\mathbf{y}}_t \in {\mathbb{C}^{M \times 1}}$ is the received signal at the BS, ${\mathbf{G}} \in {\mathbb{C}^{M \times N}}$ denotes the channel matrix between the BS and the RIS, ${\mathbf{h}_{k}} \in {\mathbb{C}^{N \times 1}}$ is the channel vector between the RIS and the UE $k$,  ${\mathbf{\Theta}}_t = {\rm{diag}}({e^{j{\varphi _1}}},{e^{j{\varphi_2}}}, \cdots ,{e^{j{\varphi_N}}})$ is the reflection coefficient matrix of the RIS, and ${\varphi_l}$ is the phase of the $l$-th RIS element. ${\mathbf{n}}_t \in {\mathbb{C}^{M \times 1}}$ is the additive white Gaussian noise (AWGN), with ${{\bf{n}}_t} \sim \mathcal{CN}( {0,{\sigma ^2}{{\bf{I}}_M}} )$. ${{{{x}}_{k,t}}}$ is the transmitted signal from the  UE $k$.
	
	In practice, the BS and the RIS are placed at the fixed positions in most situations, and UEs are mobile. Therefore, the BS-RIS channel can be regarded as quasi-static in a large timescale and the RIS-UE channels are time-varying in a small timescale. This is called the two-timescale property of channels. We define the coherence time of the large-timescale channel and the small-timescale channel as ${T_L}$ and ${T_S}$, respectively, and set ${T_L} = \tau {T_S}$. Besides, we also define the time step consists of ${T_S}$ time slots. The received signal in the $s$-th time step can be expressed as 
	\begin{align}   \label{xu2}
	{\bf{Y}}( s ) &= \sum\limits_{k = 1}^K {{\mathbf{G}}{\rm{diag}}({\mathbf{h}_{k}}(s)){\mathbf{v}}(s) {{\bf{x}}_k}  + {\bf{N}}( s )}\notag\\
	&= \sum\limits_{k = 1}^K {{{{{\mathbf{H}}_{k}}(s)}}{\mathbf{v}}(s) {{\bf{x}}_k} + {\mathbf{N}}(s)},
	\end{align}
	where ${\bf{Y}}( s )\in {\mathbb{C}^{M \times T_S}}$ is the received signal in the $s$-th time step, ${\mathbf{h}_{k}}(s) \in {\mathbb{C}^{N \times 1}}$ is the channel vector between the RIS and the UE $k$ in the $s$-th time step, ${{{{\mathbf{H}}_{k}(s)}} \buildrel \Delta \over ={\mathbf{G}}{\rm{diag}}({{{\mathbf{h}}_{k}(s)}})}\in {\mathbb{C}^{M \times N}}$ denotes the BS-RIS-UE cascaded channel, ${\mathbf{v} }(s) \buildrel \Delta \over = [{e^{j{\varphi_1}}},{e^{j{\varphi_2}}}, \cdots ,{e^{j{\varphi_N}}}]^T \in {\mathbb{C}^{N \times 1}}$ is the reflection coefficient vector of the RIS, ${{\bf{x}}_k} = [ {{x_{k,1}},{x_{k,2}}, \cdots ,{x_{k,T_S}}} ] \in {\mathbb{C}^{1 \times T_S}}$ is the transmitted signal sequence, and ${\mathbf{N}}(s) = [{{\bf{n}}_1},{{\bf{n}}_2}, \cdots ,{{\bf{n}}_{{T_S}}}] \in {\mathbb{C}^{M \times T_S}}$ is the AWGN matrix.

	\subsection{Time-Varying mmWave Channel Model}
	Due to the two-timescale property, we model ${\mathbf{G}}$ as a mmWave channel without the Doppler effect and ${{{\mathbf{h}}_k}}$ as a time-varying mmWave channel with the Doppler effect. 
	
	The mmWave channel is well described by the Saleh-Valenzuela model \cite{wangSVmodel}. Thus ${\mathbf{G}}$ can be expressed as 
	\begin{align}   \label{xu4}
	{\mathbf{G}} = \sqrt {\frac{{MN}}{{{L_G}}}} \sum\limits_{\rho = 1}^{{L_G}} {{\alpha _\rho}} {{\mathbf{a}}_r}\left( {v _r^\rho,\phi _r^\rho} \right){\mathbf{a}}_t^H\left( {v _t^\rho,\phi _t^\rho} \right),
	\end{align}
	where $L_G$, ${\alpha _\rho }\sim {\cal C}{\cal N}(0,1)$, ${v _r^\rho}$ $({\phi _r^\rho})$, and ${v _t^\rho}$ $({\phi _t^\rho})$$(\rho  = 1,2, \cdots ,{L_G})$ denote the multi-path number of $G$, the complex gain, azimuth (elevation) angle of arrival (AoA), and azimuth (elevation) angle of departure (AoD) of the $\rho$-th path of the BS-RIS channel. ${{\mathbf{a}}_r}$ $({\mathbf{a}}_t^H)$ denote the array response vector of arrival (departure). 
	
	The time-varying geometric channel model \cite{TSVmodel} can describe ${{{\mathbf{h}}_k}}$, which is given by
	\begin{align}   \label{xu5} 
	{{\bf{h}}_{k}(s)} = \sqrt {\frac{N}{{{L_k}}}} \sum\limits_{\gamma = 1}^{{L_k}} {{\beta _{\gamma,k}}{e^{j2\pi {f_{\gamma,k}}{T_S}s}}} {\bf{a}}\left( {\psi _k^\gamma,\varphi _k^\gamma} \right),
	\end{align}
	where $L_k$, ${{\bf{h}}_{k}(s)}$, ${\beta _{\gamma,k}}\sim {\cal C}{\cal N}(0,1)$, and ${\psi _{k}^\gamma}$ $({\varphi _{k}^\gamma})$$(\gamma  = 1,2, \cdots ,{L_k})$ denote the multi-path number of ${{{\mathbf{h}}_k}}$, the time-varying channel, the complex gain, and azimuth (elevation) AoD of the $\gamma$-th path of the RIS-UE channel during $s$-th time step, respectively. ${f_{\gamma,k}}$ is the Doppler shift. The array response vector of a half-wavelength spaced uniform planar array is given by
	\begin{equation}   \label{xu6}
	\begin{array}{l}
	{\bf{a}}\left( {\theta ,\phi } \right) = \frac{1}{{\sqrt {{N_x}{N_y}} }}\left[ {1, \cdot  \cdot  \cdot ,{e^{j\pi \left( {{n_x}\sin \theta \sin\phi  + {n_y}\cos \phi } \right)}}, \cdot  \cdot  \cdot ,} \right.\\
	\;\;\;\;\;\;\;\;\;\;\;\;\;\;\;\;\;\;\;\;\;\;\;\;\;\;\;\;{\left. {{e^{j\pi \left( {\left( {{N_x} - 1} \right)\sin \theta \sin\phi  + \left( {{N_y} - 1} \right)\cos \phi } \right)}}} \right]^T},
	\end{array}
	\end{equation}
	where $N_x$ and $N_y$ are the number of horizontal and vertical antennas of the plane. $n_x$ and $n_y$ are the antenna element indices. $\theta$ and $\phi$ are the azimuth and elevation angles, respectively.	

	\section{The Proposed Two-timescale Joint Channel Decomposition and Prediction Framework}
	In this section, we propose a joint channel decomposition and prediction framework based on the two-timescale channel property. By leveraging the two-timescale property, we only need to estimate $\bf{G}$ and $ {{{\bf{h}}_k}}$ once in a long and a short coherence block, respectively.
	
	\begin{figure}[tbp]
		\centering {
			\begin{tabular}{ccc}
				\includegraphics[width=0.8\textwidth]{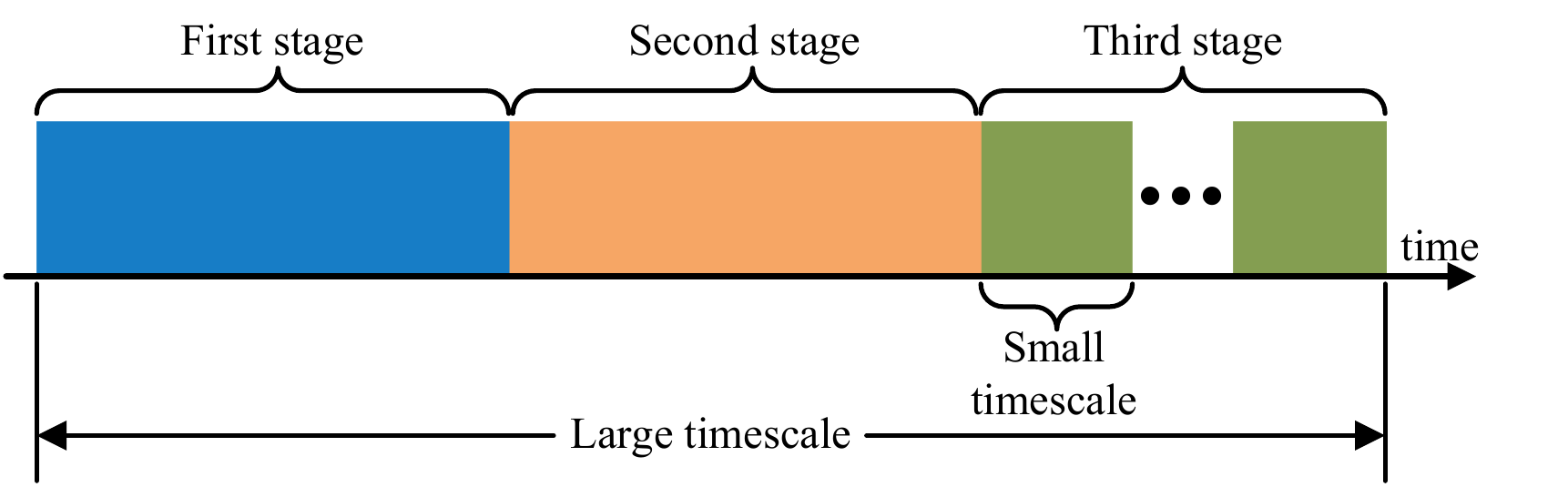}\\
			\end{tabular}
		}
		\caption{The proposed joint channel decomposition and prediction framework.}
		\vspace{-1\baselineskip}
		\label{framework}
	\end{figure}
	
	The proposed framework consists of three stages, as shown in Fig. \ref{framework}. In the first stage, we exploit the dual-link pilot transmission scheme to estimate the quasi-static channel ${{{\bf{g}}_1}}$ (${{{\bf{g}}_1}} = {\bf{G}}\left( {1,:} \right) \in {\mathbb{C}^{1 \times N}}$), thereby solving the scaling ambiguity in the third stage. We then estimate the cascaded channel of $S$ time steps in the second stage, and $S$ is the number of previous time steps needed for the channel prediction. In the third stage, we decompose $\bf{G}$ and ${ {{{\bf{h}}_k}}}$ from ${ {{{\bf{H}}_k}}}$ and finish the time-varying prediction of ${ {{{\bf{h}}_k}}}$.
	
	
	\subsection{The First Stage: The Estimation of ${{{\bf{g}}_1}}$}
	In this stage, we estimate ${{\bf{g}}_1}$ based on a dual-link pilot transmission scheme to eliminate the scale ambiguity caused by the decomposition of the third stage, which is inspired by \cite{twotime}. However, it is different that we only utilize one specific full-duplex antenna (FDA) instead of all antennas at the BS, which means less pilot overhead and hardware consumption.
	
	Specifically, a FDA at the BS transmits the pilots to the RIS via the downlink channel, and then the RIS reflects the pilots to this FDA through the uplink channel. We can exploit multiple beams to exhaustively search through the whole space to solve the problem that we cannot deliberately design one RIS beam to reflect the pilots to the direction of the FDA without channel knowledge \cite{twotime}.  We assume that the downlink channel and the uplink channel between the specific FDA and RIS are reciprocal \cite{huyixing}, represented as ${\bf{g}}_1^T$ and ${{{\bf{g}}_1}}$, respectively. The received pilots at the specific FDA can be expressed as 
	\begin{align}   \label{xu7}
	{y_t} = {{{{\bf{g}}_1}}{\rm{diag}}({\bf{v}} _t ){\bf{g}}_1^T}{x_t} + {e_t}\notag\\
	\;\;\;\; = \left( {{{\bf{g}}_1} \odot {\bf{g}}_1^T} \right){\bf{v}} _t{x_t} + {e_t},
	\end{align}
	where ${y_t}$ and ${x_t}$ are the received and transmitted pilot by the specific FDA at the $t$-th time slot, respectively. ${e_t}$ is the error resulting from all interference and noise. Therefore, we can obtain the estimated channel ${\hat{{\bf{g}}}_1}$ between the RIS and the FDA at the BS via the traditional least square (LS) method in the first stage.
	
	
	\subsection{The Second Stage: The Estimation of Cascade Channels}
	In this stage, we estimate a sequence of cascaded channels ${ {{{{\bf{\bar H}}}_k}} }$, where ${{{\bf{\bar H}}}_k} = \{ {{{\widehat {\bf{H}}}_{k}(1)},{{\widehat {\bf{H}}}_{k}(2)},}$ $ { \cdot  \cdot  \cdot ,{{\widehat {\bf{H}}}_{k}(S)}}\}$ during $S$ time steps  via the pilot-based method for the channel prediction in the third stage. The signal model is the same as (\ref{xu2}) except that ${{\bf{x}}_k} = [ {{x_{k,1}},{x_{k,2}}, \cdots ,{x_{k,K}}} ]$ denotes the orthogonal pilot to distinguish the pilots from different UEs.
	\begin{equation}   \label{new2}
	{{\bf{x}}_{{i}}}{\bf{x}}_{{j}}^H{\rm{ = }}\left\{ \begin{array}{l}
	K{P_{k}},\;\;\;\;\;{i} = {j} = {k},\\
	\;\;\;\;0,\;\;\;\;\;\;\;\;\;{i} \ne {j},
	\end{array} \right.\;\;\;\;
	\end{equation}
	where $P_{k}$ denotes the transmitted power of UE $k$.
	
	Although the estimation accuracy of ${ {{{{\bf{\bar H}}}_k}} }$ is vital for channel prediction, high-performance pilot-based channel estimation methods often require a large number of pilots. Therefore, we use the channel estimation algorithm proposed in \cite{twotime} to achieve a tradeoff between accuracy and pilot overhead. 

	\subsection{The Third Stage: The Channel Decomposition and Prediction}
	The channel decomposition and prediction at this stage does not need to transmit any pilots and makes up for the shortcomings of traditional channel estimation methods with poor real-time performance when the channel is time-varying. 
	
	%
	
	As we mentioned earlier, the RIS-assisted system has the two-timescale property, only ${ {{{\bf{h}}_k}}}$ is time-varying in a small timescale, while ${\bf{G}}$ is quasi-static in a large timescale. Therefore, we only need to perform channel prediction on ${ {{{\bf{h}}_k}}}$. However, it is challenging to obtain ${\bf{G}}$ and ${ {{{\bf{h}}_k}}}$, respectively, due to the passive elements at the RIS. Most of the existing works major in the study of cascaded channel estimation. At the same time, some works that estimate ${\bf{G}}$ and ${ {{{\bf{h}}_k}}}$ tend to arrange active elements with receiving capability at the RIS, which causes additional hardware consumption. Another feasible way to solve this problem is the decompose-based algorithms.
	
	In this stage, we propose a novel NN architecture, SCLSTM, to joint decompose  ${\bf{G}}$ and ${ {{{\bf{h}}_k}}}$ from ${ {{{\bf{H}}_k}}}$ and capture the temporal relationship of ${ {{{\bf{h}}_k}}}$ to finish the channel prediction. The details of the SCLSTM and the corresponding algorithm are introduced in Section IV.

	\section{The Proposed SCLSTM-based Algorithm for Joint Channel Decomposition and Prediction Framework}
	In this section, we first introduce the LSTM, which is the fundamental component of SCLSTM. Then, we propose two types of sparse-connected layers to decompose ${\bf{G}}$ and ${ {{{\bf{h}}_k}}}$ from ${ {{{\bf{H}}_k}}}$ based on the mapping relationship between input and output. Finally, we introduce the channel prediction process based on LSTM and the joint training of the channel decomposition and prediction.
	\subsection{The Introduction of LSTM}
	The LSTM is proposed as a special kind of recurrent neural network based on sequence operations \cite{LSTM1}. It can build up the temporal correlations between previous information and the current circumstances by passing the decision at the previous time step $s-1$ to the later time step $s$ for the time series problem. 
	
\begin{figure}[tbp]
		\centering {
			\begin{tabular}{ccc}
				\includegraphics[width=0.8\textwidth]{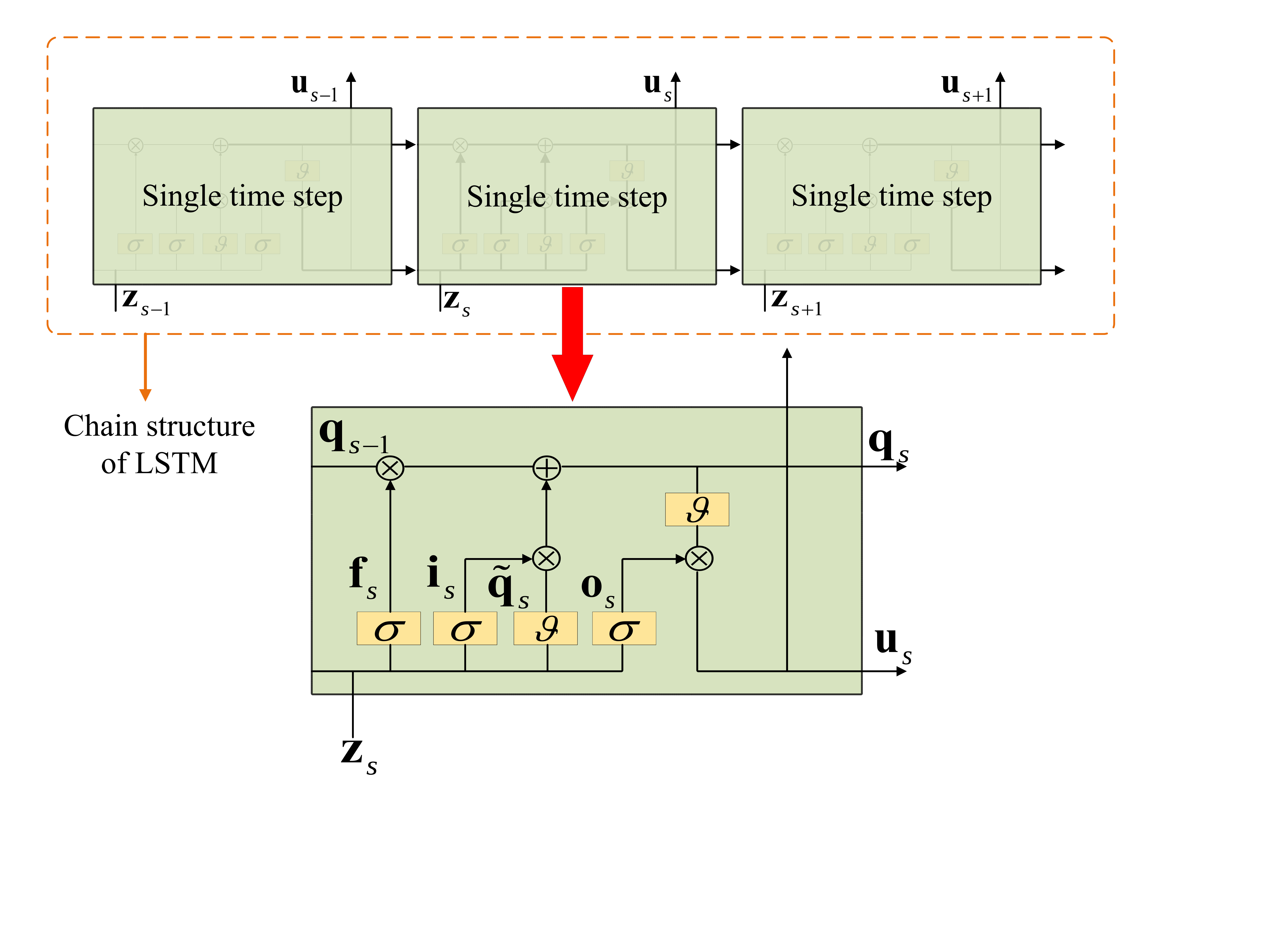}\\
			\end{tabular}
		}
		\caption{The structure of the LSTM network.}
		\vspace{-1\baselineskip}
		\label{lstm}
	\end{figure}

	The details of the LSTM structure are shown in Fig. \ref{lstm}. We generally define the input sequence of the LSTM as $\left\{ {{{\bf{z}}_1},{{\bf{z}}_2},...,{{\bf{z}}_S}} \right\}$, and ${{\bf{z}}_s} \in {\mathbb{R}}{^F}$, where $F$ is the dimension of the input data in each time step. The key of establishing temporal connections in the LSTM is the internal memory cell states defined as $\left\{ {{{\bf{q}}_1},{{\bf{q}}_2},...,{{\bf{q}}_S}} \right\}$ and maintained throughout the cycle, which are the most critical elements of the LSTM structure. The memory cell state runs straight down the entire chain and interacts with the intermediate output and the next time step input to determine which elements of the internal state vector should be updated, maintained, or erased. Note that the information of the memory cell state is regulated by the structure called the gate, which is illustrated in Fig. \ref{lstm}. The gate is composed out of a sigmoid layer and a pointwise multiplication operation. The output of the sigmoid layer is between 0 and 1, describing how much information each component should be allowed, which 0 means no pass and 1 means all pass. There are three gates to protect and control the cell state in the LSTM. 
	
	Next, we introduce the main steps of the LSTM. At the first step, the LSTM uses a sigmoid layer called the ``forget gate layer'' to throw away some information from the memory cell state selectively. As shown in Fig. \ref{lstm}, the inputs of the ``forget gate layer'' are the intermediate output ${{\bf{u}}_{s - 1}}$ at the previous time step and the input ${{\bf{z}}_s}$ at the current time step, and the output ${{\bf{f}}_s}$ is a real value vector whose elements are between 0 and 1 for each component of the memory cell state ${{\bf{q}}_{s-1}}$. The operation of the first step can be expressed as 
	\begin{align}   \label{xu8}
	{{\bf{f}}_s} = \sigma \left( {{{\bf{W}}_{fz}}{{\bf{z}}_s} + {{\bf{W}}_{fu}}{{\bf{u}}_{s-1}} + {{\bf{b}}_f}} \right),
	\end{align}
	where  ${{{\bf{W}}_{fz}}}$ and ${{{\bf{W}}_{fu}}}$ are weight matrices for the corresponding inputs of the network activation functions, respectively. ${{\bf{b}}_f}$ is the bias vector. $\sigma$ is the sigmoid activation function defined as
	\begin{align}   \label{xu14}
	\sigma \left( x \right) = \frac{1}{{1 + {e^{ - x}}}}.
	\end{align}
	
	The second step is to store new information in the memory cell state, and it consists of two parts. The first part is to apply a tanh layer to generate a vector of new candidate values, ${{{\bf{\tilde q}}}_s}$, which would be added to the memory cell state. In the second part, the output ${{\bf{i}}_s}$ of a sigmoid layer called the ``input gate layer'' is applied to choose the values of ${{{\bf{q}}}_s}$ that need to be updated. Then, the current state ${{{\bf{q}}}_s}$ can be updated as 
	\begin{align}   
	{{{\bf{\tilde q}}}_s} &= \vartheta \left( {{{\bf{W}}_{qz}}{{\bf{z}}_s} + {{\bf{W}}_{qu}}{{\bf{u}}_{s - 1}} + {{\bf{b}}_q}} \right),\label{xu9}\\
	{{\bf{i}}_s} &= \sigma \left( {{{\bf{W}}_{iz}}{{\bf{z}}_s} + {{\bf{W}}_{iu}}{{\bf{u}}_{s-1}} + {{\bf{b}}_i}} \right), \label{xu10}\\
	{{\bf{q}}_s} &= {{{\bf{\tilde q}}}_s} \odot {{\bf{i}}_s} + {{\bf{q}}_{s - 1}} \odot {{\bf{f}}_s}, \label{xu11}
	\end{align} 
	
	where ${{{\bf{W}}_{qz}}}$, ${{{\bf{W}}_{qu}}}$, ${{{\bf{W}}_{iz}}}$, and ${{{\bf{W}}_{iu}}}$ are weight matrices for the corresponding inputs of the network activation functions, respectively. ${{\bf{b}}_i}$ and ${{\bf{b}}_q}$ are the bias vectors. $\vartheta $ denotes the tanh activation function defined as 
	\begin{align}   \label{xu15}
	\vartheta \left( x \right) = \frac{{{e^{2x}} - 1}}{{{e^{2x}} + 1}}.
	\end{align}
	
	As we can see from (\ref{xu9})-(\ref{xu11}), the current state ${{\bf{q}}_s}$ is determined by the state preserved in the previous time step and the newly added state.
	
	In the third step, the output of the LSTM block ${{\bf{u}}_{s}}$ is determined by the ${{\bf{q}}_s}$, ${{\bf{u}}_{s-1}}$, and ${{\bf{z}}_s}$. We first use a sigmoid layer to decide which parts of the ${{\bf{q}}_s}$ need to be output. Then, we limit the values of the ${{\bf{q}}_s}$ between -1 and 1 by a tanh function and multiply them by the output ${{\bf{o}}_s}$ of the sigmoid gate to output the parts we need to.
	\begin{align}   
	{{\bf{o}}_s} &= \sigma \left( {{{\bf{W}}_{oz}}{{\bf{z}}_s} + {{\bf{W}}_{ou}}{{\bf{u}}_{s - 1}} + {{\bf{b}}_o}} \right),\label{xu12}\\
	{{\bf{u}}_s} &= \vartheta \left( {{{\bf{q}}_s}} \right) \odot {{\bf{o}}_s}, \label{xu13}
	\end{align}
	where ${{{\bf{W}}_{oz}}}$ and ${{{\bf{W}}_{ou}}}$ are weight matrices for the corresponding inputs of the network activation functions. ${{\bf{b}}_o}$ is the bias vector. 
	
	This three-step process continues to repeat many time steps to build a chain structure of LSTM modules. The details of the chain structure are shown in Fig. \ref{lstm}. It can selectively preserve and maintain the information of the current time step, which can affect the future time steps. 
	\subsection{The Joint Channel Decomposition and Prediction Algorithm Based on the Proposed SCLSTM}
	\begin{figure*}[tbp]
		\centering {
			\begin{tabular}{ccc}
				\includegraphics[width=1\textwidth]{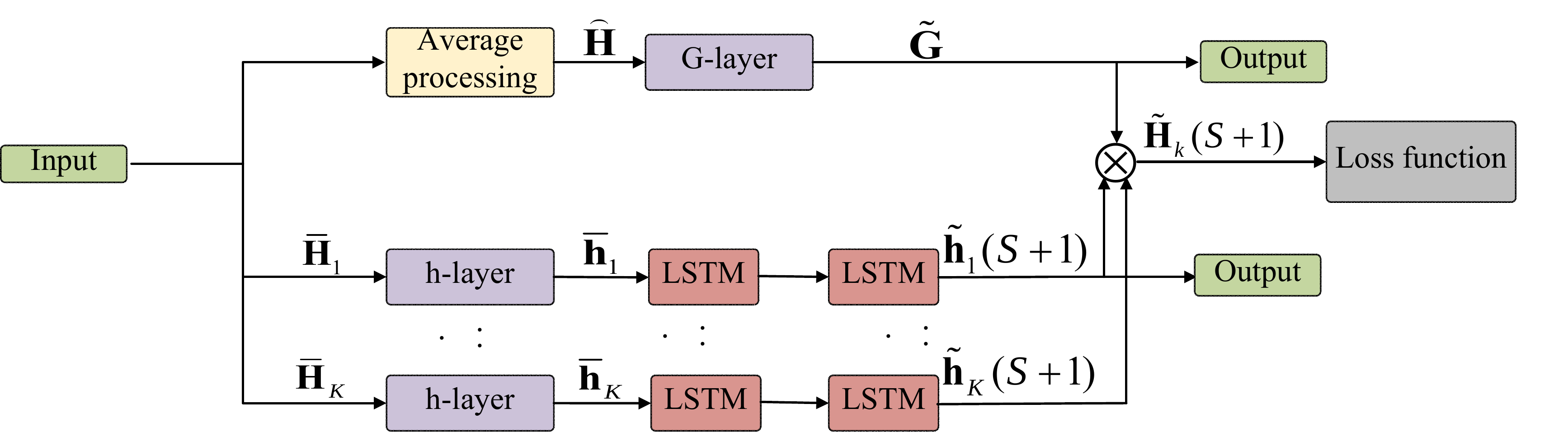}\\
			\end{tabular}
		}
		\caption{The structure of SCLSTM network.}
		\vspace{-1\baselineskip}
		\label{sclstm}
	\end{figure*}
	The structure of our proposed SCLSTM is shown in Fig. \ref{sclstm}, which comprises one sparse-connected layer called G-layer, $K$ sparse-connected layer called h-layer, and $K$ LSTM blocks. Each LSTM block consists of two LSTMs. For $K$ different UEs, all h-layers share the same set of parameters, and so do all LSTM blocks. Therefore, for convenience, we mainly introduce the joint decomposition and prediction of the UE $k$, which also applies to multiple UEs. The input of the SCLSTM is the ${{{{{\bf{\bar H}}}_k}}}$ estimated in the second stage. Besides, the SCLSTM is trained based on real values, so we stack the real and imaginary parts of the ${\bf{\bar H}}_k$ into a vector which can be defined as
	\begin{align}   \label{xu16}
	\begin{array}{l}
	\rm{Input} = \left[ {\rm{vec}\left( {Re\left\{ {{{{\bf{\bar H}}}_1}} \right\}} \right),\rm{vec}\left( {{\mathop{\rm Im}\nolimits} \left\{ {{{{\bf{\bar H}}}_1}} \right\}} \right), \cdots ,} \right.\\
	\;\;\;\;\;\;\;\;\;\;\;\;\;\;\;{\left. {{\rm{vec}}\left( {{\rm{Re}}\left\{ {{\bf{\bar H}}}_K \right\}} \right),{\rm{vec}}\left( {{\mathop{\rm Im}\nolimits} \left\{ {{{{\bf{\bar H}}}_K}} \right\}} \right)} \right]^T}.
	\end{array}
	\end{align}
	The outputs of the SCLSTM are ${{\bf{\tilde G}}}$ and ${{{\bf{\tilde h}}}_k}(S + 1)$.
	
	For the UE $k$, we aim to decompose ${\bf{G}}$ and ${{\bf{h}}_k}$ from ${\bf{H}}_k$ in the channel decomposition part. As we defined above, ${{{{\mathbf{H}}_k}}}={\mathbf{G}}{\rm{diag}}({{{\mathbf{h}}_k}})$, ${\bf{H}}_k$ is the matrix product of ${\bf{G}}$ and ${\bf{h}}_k$, which means that we can obtain ${\bf{G}}$ and ${\bf{h}}_k$ by matrix operation on ${\bf{H}}_k$. However, it is tough to implement this matrix operation without prior knowledge about ${\bf{G}}$ and ${\bf{h}}_k$, which are hard to obtain because of the passive elements on the RIS that cannot receive and process signals.
	
	To overcome this dilemma, we can exploit the fully-connected layer to replace the decomposition matrix operations unlike some existing work based on complex PARAFAC decomposition. The operation of the original fully-connected layer is similar to matrix operation, which can be expressed as 
	\begin{align}   \label{xu17}
	{{\bf{x}}_{out}} = \alpha \left( {{\bf{W}}{{\bf{x}}_{in}} + {\bf{b}}} \right),
	\end{align}
	where ${\bf{W}}$ and ${\bf{b}}$ are the weights matrix and the bias vector that can be learned, respectively; $\alpha$ is the activation function of the layer. However, each element of the output of the fully-connected layer has a corresponding connection with each element of the input. To reduce the number of connections, we propose two types of sparse-connected, called G-layer and h-layer, to decompose ${\bf{G}}$ and ${{\bf{h}}_k}$ with less computational complexity directly. To ensure that ${\bf{G}}$ is constant over $S$ time steps, we first perform the following average processing on the input of the G-layer while the input of the $k$-th h-layer is ${\rm{Inpu}}{{\rm{t}}_{h,k}} = {[ {{\rm{vec}}( {{\rm{Re}}\{ {{{{\bf{\bar H}}}_k}} \}} ),{\rm{vec}}( {{\rm{Im}}\{ {{{{\bf{\bar H}}}_k}} \}} )} ]^T}$.
	\begin{align}   \label{xu18}
	{\bf{\mathord{\buildrel{\lower3pt\hbox{$\scriptscriptstyle\frown$}} 
				\over H} }} = \frac{{{{{\bf{\hat H}}}_{1}(1)}  +  \cdot  \cdot  \cdot  + {{{\bf{\hat H}}}_{1}(S)} + {{{\bf{\hat H}}}_{2}(1)} +  \cdot  \cdot  \cdot  + {{{\bf{\hat H}}}_{K}(S)}}}{{SK}},
	\end{align}
	\begin{align}   \label{xu19}
	{\rm{Inpu}}{{\rm{t}}_G} = {[ {{\mathop{\rm Re}\nolimits} \{ {{\bf{\mathord{\buildrel{\lower3pt\hbox{$\scriptscriptstyle\frown$}} 
							\over H} }}} \},{\mathop{\rm Im}\nolimits} \{ {{\bf{\mathord{\buildrel{\lower3pt\hbox{$\scriptscriptstyle\frown$}} 
							\over H} }}} \}} ]^T} \in {\mathbb{R}^{2MN \times 1}}.
	\end{align}
	
	Besides, the outputs of the two types of sparse-connected layers are also different, while one is ${{\bf{\tilde G}}}$ and the other is ${{\bf{\bar h}}_k} = \{ {{{{\bf{\hat h}}}_{k}(1)},{{{\bf{\hat h}}}_{k}(2)},...,{{{\bf{\hat h}}}_{k}(S)}} \}$. 
	
\begin{figure}[tbp]
		\centering {
			\begin{tabular}{ccc}
				\includegraphics[width=0.7\textwidth]{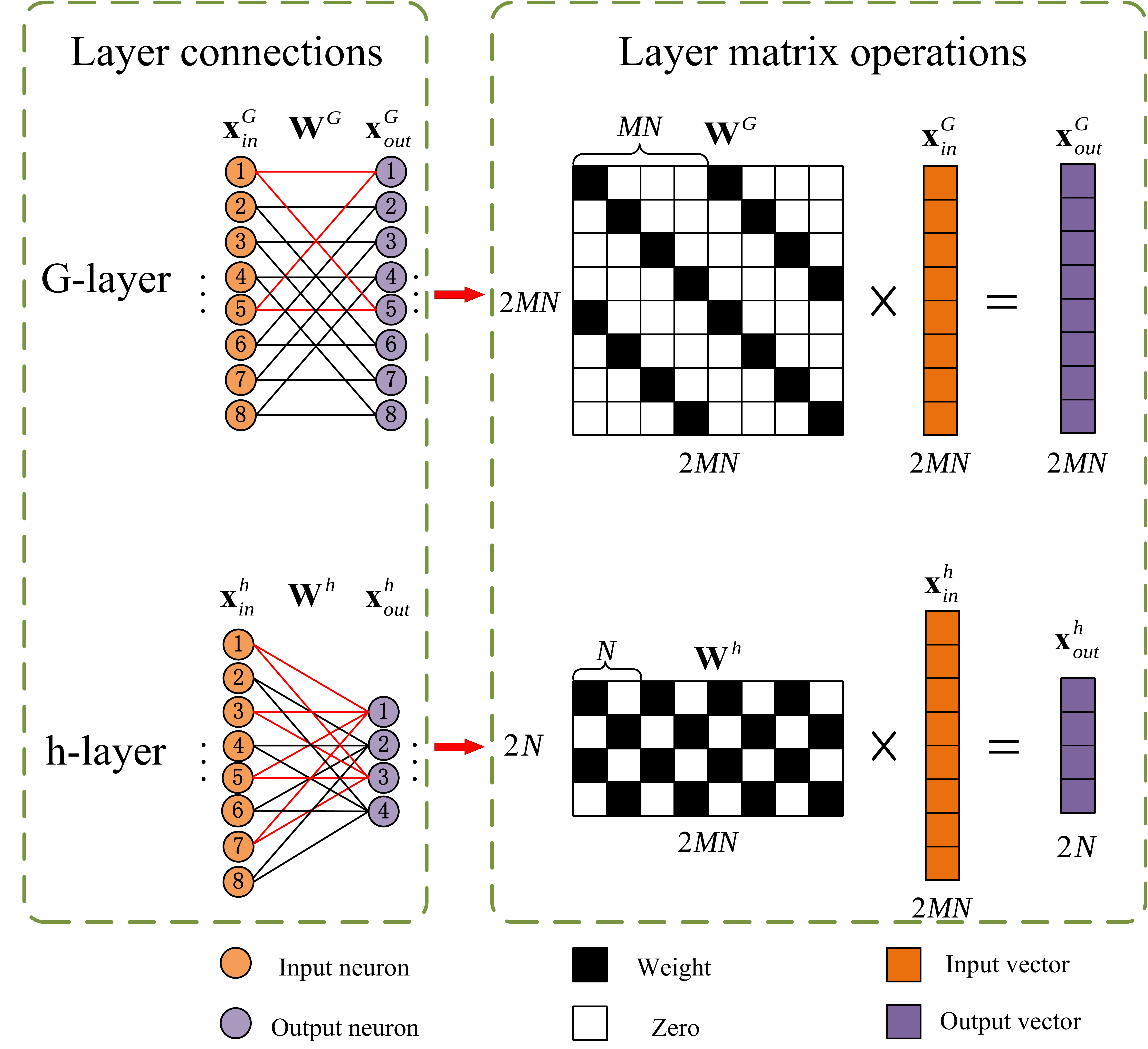}\\
			\end{tabular}
		}
		\caption{Schematic diagram of two sparse-connected layers.}
		\vspace{-1\baselineskip}
		\label{sparselayer}
	\end{figure}

	According to the definition of the cascaded channel, the input and output formats of the two sparse layers, we can find that the real and imaginary parts of ${\bf{G}}$ or ${{\bf{h}}_k}$ in the output of the two sparse layers are only related to the inputs with the information of ${\bf{G}}$ or ${{\bf{h}}_k}$, respectively. The elements ${\mathop{\rm Re}\nolimits} \{ {{{\bf{G}}{[ {m,n} ]}}} \}$ and ${\mathop{\rm Im}\nolimits} \{ {{{\bf{G}}{[ {m,n} ]}}} \}$ of the G-layer's output are determined by the elements ${\mathop{\rm Re}\nolimits} \{ {{{\bf{G}}{[ {m,n} ]}}{{\bf{h}}_{k}{[ {n} ]}}} \}$ and ${\mathop{\rm Im}\nolimits} \{ {{{\bf{G}}{[ {m,n} ]}}{{\bf{h}}_{k}{[ {n} ]}}} \}$ in the input, and ${\mathop{\rm Re}\nolimits} \{ {{{\bf{h}}_{k}{[ {n} ]}}} \}$ and ${\mathop{\rm Im}\nolimits} \{ {{{\bf{h}}_{k}{[ {n} ]}}} \}$ of the h-layer's output are determined by the ${\mathop{\rm Re}\nolimits} \{ {{{\bf{G}}{{[ {1:M},n ]}}}{{\bf{h}}_{k}{[ {n} ]}}} \}$ and ${\mathop{\rm Im}\nolimits} \{ {{{\bf{G}}{{[ {1:M},n ]}}}{{\bf{h}}_{k}{[ {n} ]}}} \}$. Therefore, we can use this property to build two sparse-connected layers, as shown in Fig. \ref{sparselayer}. In Fig. \ref{sparselayer}, the layer connections and corresponding layer matrix operations are illustrated in detail. For example, we assume that $M=N=2$. The output neurons ${\bf{x}}_{out}^G[1]$ and ${\bf{x}}_{out}^G[5]$ of the G-layer represent ${\mathop{\rm Re}\nolimits} \{ {{{\bf{G}}{[ {1,1} ]}}} \}$ and ${\mathop{\rm Im}\nolimits} \{ {{{\bf{G}}{[ {1,1} ]}}} \}$, respectively. They only connect to ${\mathop{\rm Re}\nolimits} \{ {{{\bf{G}}{[ {1,1} ]}}{{\bf{h}}_{k}{[ {1} ]}}} \}$ and ${\mathop{\rm Im}\nolimits} \{ {{{\bf{G}}{[ {1,1} ]}}{{\bf{h}}_{k}{[ {1} ]}}} \}$ marked as ${\bf{x}}_{in}^G[1]$ and ${\bf{x}}_{in }^G[5]$. Similarly, the output neurons ${\bf{x}}_{out}^h[1]$ and ${\bf{x}}_{out}^h[3]$ of the h-layer represent ${\mathop{\rm Re}\nolimits} \{ {{{\bf{h}}_{k}{[ {1} ]}}} \}$ and ${\mathop{\rm Im}\nolimits} \{ {{{\bf{h}}_{k}{[ {1} ]}}} \}$, respectively. They both have connections with ${\mathop{\rm Re}\nolimits} \{ {{{\bf{G}}{[ {1,1} ]}}{{\bf{h}}_{k}{[ {1} ]}}} \}$, ${\mathop{\rm Re}\nolimits} \{ {{{\bf{G}}{[ {2,1} ]}}{{\bf{h}}_{k}{[ {1} ]}}} \}$, ${\mathop{\rm Im}\nolimits} \{ {{{\bf{G}}{[ {1,1} ]}}{{\bf{h}}_{k}{[ {1} ]}}} \}$ and ${\mathop{\rm Im}\nolimits} \{ {{{\bf{G}}{[ {2,1} ]}}{{\bf{h}}_{k}{[ {1} ]}}} \}$ marked as ${\bf{x}}_{in}^h[1]$, ${\bf{x}}_{in}^h[3]$, ${\bf{x}}_{in}^h[5]$, and ${\bf{x}}_{in}^h[7]$, respectively. Therefore, both weight matrixes ${\bf{W}}^G$ and ${\bf{W}}^h$ of two sparse-connected layers are sparse matrixes with $4MN$ entries, while the corresponding fully-connected weight matrixes have $4{M^2}{N^2}$ and $4{M}{N^2}$ entries, respectively.
	
	
	Next, we utilize the decomposed sequence ${{\bf{\bar h}}_k}$ as the input of the $k$-th LSTM block to predict the ${{{{\bf{\tilde h}}}_{k}(S + 1)}}$ at the $S+1$ time step. The LSTM block consists of two LSTMs with $6N$ and $4N$ neurons, and one dense layer with $2N$ neurons. Thus, the SCLSTM can adapt to the changes in the size of the RIS. Note that the output of the dense layer is the vector of the real and imaginary parts of ${{{{\bf{\tilde h}}}_{k}(S + 1)}}$. Then, we need to transform it into a complex vector.   
	
	To train the SCLSTM for joint channel decomposition and prediction, we first calculate the predicted cascaded channel ${{{\bf{\tilde H}}}_{k}(S + 1)}$ using the decomposed ${{{\bf{\tilde G}}}}$ and the predicted ${{{\bf{\tilde h}}}_{k}(S + 1)}$, and then use the mean squared error (MSE) between the ${{{\bf{\tilde H}}}_{k}(S + 1)}$ and its desired values as the loss function of SCLSTM, which can be expressed as  
	\begin{align}   \label{xu20}
	{\rm{MSE}} = \frac{1}{B}\sum\limits_{b = 1}^B {\sum\limits_{k = 1}^K {{{\left\| {{{{\bf{\tilde H}}}_{k}(S + 1)} - {{\bf{H}}_{k}(S + 1)}} \right\|}^2}} },
	\end{align}
	where $B$ is the size of a training batch, and ${{{\bf{H}}_{{{k}}}(S + 1)}}$ is the desired cascaded channel at the $S+1$ time step. As we can see from (\ref{xu20}), the loss function includes the parameters of the two sparse-connected layers for channel decomposition and the LSTM structure for channel prediction, so their parameters can be updated simultaneously to achieve the joint training. 
	
	Similar to PARAFAC based methods in \cite{CEhuang}, \cite{chidu2}, ${{{\bf{\tilde G}}}}$ and ${{{\bf{\tilde h}}}_{k}(S + 1)}$ decomposed based on SCLSTM algorithm also have scaling ambiguity, which can be expressed as
	\begin{align}   \label{xu21}
	{\bf{\tilde G}} = {\bf{G}}{\bf{\Delta}} _{1},\;{\rm{diag}}({{{\bf{\tilde h}}}_{k}(S + 1)}) = {{\bf{\Delta}} _{2}}{\rm{diag(}}{{\bf{h}}_{k}({S + 1})}),
	\end{align}
	where ${\bf{\Delta}} _{1}$ and ${\bf{\Delta}} _{2}$ are $N \times N$ complex diagonal scaling matrices, and ${{\bf{\Delta }}_1}{{\bf{\Delta }}_{\rm{2}}}{\rm{ = }}{{\bf{I}}_N}$. ${{{\bf{g}}_1}}$ is used to eliminate the scaling ambiguity by correcting the first row of ${\bf{\tilde G}}$. 
	
	After the training is completed, we reserve the model with the best performance on the validation set. For continuous online prediction, we first obtain $\hat{{\bf{g}}}_1$ and ${ {{{{\bf{\bar H}}}_k}} }$ in the first and second stages, respectively. Secondly, they are used to predict ${{{\bf{\tilde H}}}_k}(S + 1)$ at the ($S+1$)-th time step. Then we use ${{{\bf{\bar H}}}_k} = [{\widehat {\bf{H}}_k}(2),{\widehat {\bf{H}}_k}(3), \cdot  \cdot  \cdot ,{{{\bf{\hat H}}}_k}(S + 1)]$ to predict ${{{\bf{\tilde H}}}_k}(S + 2)$, and so on until the large-timescale ${{{\bf{G}}}}$ changes and the first and second stages need to be repeated. Finally, we use $\hat{{\bf{g}}}_1$ to eliminate the scaling ambiguity. The details of the joint SCLSTM-based channel decomposition and prediction algorithm are shown in Table \ref{algorithm}.
	\begin{table}[tbp]\normalsize
		\centering
		\caption{\label{algorithm}} 
		\begin{tabular}{lcl} 
			\toprule 
			\textbf{Algorithm 1: A SCLSTM-based algorithm} \\ 
			\midrule 
			\textbf{Offline Training:}\\
			\quad\textbf{Input:} Training set $\{ {{{{\bf{\bar H}}}_k},{{\bf{H}}_{k}(S + 1)}} \}$.\\
			\quad\textbf{Output:} Trained SCLSTM model. \\
			1.\quad Obtain the input ${{\bf{\mathord{\buildrel{\lower3pt\hbox{$\scriptscriptstyle\frown$}} 
							\over H} }}}$ of G-layer by (\ref{xu18}); \\
			2.\quad Obtain the output ${{\bf{\tilde G}}}$ of G-layer; \\ 
			3.\quad Obtain the output ${{{\bf{\bar h}}}_k}$ of h-layer;\\
			4.\quad Obtain the output ${{{\bf{\tilde h}}}_k}(S + 1)$ of LSTM block;\\ 
			5.\quad Calculate  ${{{\bf{\tilde H}}}_k}(S + 1) = {\bf{\tilde G}}{\rm{diag}}({{{\bf{\tilde h}}}}_{k}(S + 1))$;\\
			6.\quad Calculate the loss function by (\ref{xu20});\\
			7.\quad Update the parameters of the SCLSTM;\\
			8.\quad Output the best SCLSTM model via a validation set;\\
			\textbf{Online Continuous Prediction:}\\
			\quad\textbf{Input:} The trained SCLSTM, the coherence time ${T_L}$ of ${\bf{G}}$, and the time period ${T_C}$; \\
			\quad\textbf{Output:} $\{ {{{{\bf{\tilde H}}}_{k}(S + 1)},...,{{{\bf{\tilde H}}}_{{k}}(T_C)}} \}$, $\{ {{{{\bf{\tilde G}}}_1},...,{{{\bf{\tilde G}}}_{\lceil {\frac{{{T_C}}}{{{T_L}}}} \rceil }}} \}$, and $\{ {{{{\bf{\tilde h}}}_{k}(S + 1)},...,{{{\bf{\tilde h}}}_{{k}}(T_C)}} \}$; \\
			9. \textbf{Initialization:} $t=S+1$, $m=S+1$;\\
			10. \textbf{If} ${T_C} \le {T_L}$\\
			11. \quad Obtain ${{{\bf{\hat g}}}_{\lceil {\frac{t}{{{T_L}}}} \rceil ,1}}$ in the first stage; \\
			12. \quad Obtain ${\bf{\bar H}}_k^{(t)} = [{{{\bf{\hat H}}}_k}(t - S),...,{{{\bf{\hat H}}}_k}(t - 1)]$; \\
			13. \quad \textbf{while} $t \le {T_C}$ \textbf{do}\\
			14. \qquad Obtain $\{ {{{\bf{\tilde H}}}_k}(t),{{{\bf{\tilde G}}}_{\lceil {\frac{t}{{{T_L}}}} \rceil }},{{{\bf{\tilde h}}}_k}(t)\} $ via the SCLSTM;\\
			15. \qquad $t=t+1$;\\  
			16. \qquad update ${\bf{\bar H}}_k^{(t)} = [{{{\bf{\hat H}}}_k}(t - S),...,{{{\bf{\hat H}}}_k}(t - 1)]$; \\
			17. \quad \textbf{end while and output}\\
			18. \textbf{Elif} ${T_C} > {T_L}$\\
			19. \quad do step $11 \sim 12$;\\
			20. \quad \textbf{while} $t \le {T_C}$ and $m \le {T_L}$ \textbf{do}\\
			21. \qquad do step $14 \sim 16$ and $m=m+1$; \\ 
			22. \qquad \textbf{if} $m  >  {T_L}$\\
			23. \qquad \quad $m = m - {T_L}$ and do step $11 \sim 12$;\\
			24. \quad \textbf{end while and output}\\
			25. Eliminate the scaling ambiguity of the outputs via $\hat{{\bf{g}}}_1$.\\
			\bottomrule 
		\end{tabular} 
	\end{table} 

	\section{Pilot Overhead and Computational Complexity}
	\subsection{Pilot Overhead}
	The pilot overhead of our proposed three-stage framework mainly comes from the first and second stages. In the first stage, we need $N$ time slots to transmit pilots for efficient estimation of ${{{\bf{g}}_1}}$ consisting of $N$ coefficients. In the second stage, there are $KSMN$ coefficients in the ${ {{{{\bf{\bar H}}}_k}} }$ that need to be estimated. Therefore, the applied channel estimation method \cite{twotime} costs ${2(N+1) + KS\lceil {\frac{N}{M}} \rceil }$ pilots, and the total pilot $P_L$ during $T_L$ is ${3N+2 + KS\lceil {\frac{N}{M}} \rceil }$.
	
	In our proposed framework, there is no pilot consumption in the channel prediction stage. Therefore, we have to average the pilot overhead to the whole framework time. We also define the average pilot overhead $P_a$ during a period of ${T_S}$, which can be expressed as
	\begin{align}   \label{new21}
	{{P_a} = \frac{ {P_L}{T_S}}{T_L} = \frac{ {P_L}} {\tau}}.
	\end{align}
	
	We then compare the value of $P_a$ of our proposed framework with some existing channel estimation methods illustrated in Table \ref{Pilot}. 
	
	\emph{\textbf{Proposition 1:}} The average pilot overhead $P_a$ of our proposed framework is lower than the MVU algorithm \cite{MVU} and PARAFAC-VAMP algorithm \cite{CEhuang} with the feasible condition $\tau  > \frac{3}{K} + \frac{S}{M} + \frac{S}{N} + \frac{2}{{NK}}$.
	
	%
	
	\emph{\textbf{Proof:}} Please refer to Appendix A.
	
	\emph{\textbf{Proposition 2:}} The average pilot overhead of our proposed framework is lower than the two-timescale framework \cite{twotime} with the feasible condition $\tau  \ge \frac{M}{K} + S$.
	
	
	\emph{\textbf{Proof:}} The proof is similar to that of proposition 1. We can get 
	\begin{align}   \label{proof21}
	\tau  \ge \frac{N}{{K\frac{N}{M}}} + S = \frac{M}{K} + S \Rightarrow \tau  \ge \frac{N}{{K\lceil {\frac{N}{M}} \rceil }} + S,
	\end{align}
	where $\lceil {\frac{N}{M}} \rceil  \ge \frac{N}{M}$ is exploited in the above derivation.
	
	The proposition 1 shows that the feasible conditions of $\tau$ for pilot comparison between the proposed framework and the algorithms in \cite{MVU} and \cite{CEhuang} are the same and determined by $K$, $M$, $S$, and $N$. When given $K$, $M$, and $S$, the lower bound of $\tau$ decreases with the increasing $N$. Besides, we can conclude that the lower bound of $\tau$ of the pilot overhead comparison between the proposed framework and the two-timescale algorithm in \cite{twotime} has nothing to do with $N$, but $K$, $M$, and $S$  from proposition 2. When $K$ and $M$ are fixed, the lower bound is only determined by $S$. In practice, $S$ is small and $\tau  >  > 1$, which means our proposed framework is significant to the pilot reduction compared to the two-timescale algorithm in \cite{twotime}.
	\begin{table*}[tbp]
		\centering
		\caption{{Pilot overhead and computational complexity comparisons of} \protect \\ {different channel estimation schemes.}}
		\begin{tabular}{|c|c|c|}
			\hline
			\vspace{-0.25cm}
			& & \\
			Channel estimation algorithms &  Average pilot overhead $P_a$ & Computational complexity\\
			\hline
			\vspace{-0.25cm}
			& & \\
			Proposed framework & $\frac{{3N + 2 + KS\lceil {\frac{N}{M}} \rceil }}{\tau }$ & ${\cal O}\left( {360K{N^2} + ( {4KM+4M + 42K} )N} \right)$\\
			\hline
			\vspace{-0.25cm}
			& & \\
			MVU \cite{MVU}  & $NK$ & ${\cal O}( {{N^3} + K{N^2}} )$  \\
			\hline
			\vspace{-0.25cm}
			& & \\
			\makecell[c]{ PARAFAC-VAMP\cite{CEhuang} } &  $KMP$ & ${\cal O}( {( {K + M} )( {5{N^2} - N} )} )$ \\
			\hline
			\vspace{-0.25cm}
			& & \\
			\makecell[c]{Two-Timescale \cite{twotime}} & $\frac{{2( {N + 1} )}}{\tau} + {K\lceil {\frac{N}{M}} \rceil}$ & ${\cal O}( {2{N^3} + ( {K + ML} ){N^2} + MLN{I_{\max }}} )$\\
			\hline
		\end{tabular}
		\label{Pilot}
	\end{table*}
	\subsection{Computational Complexity}
	In Table \ref{Pilot}, we summarize the computational complexity of our proposed SCLSTM and other traditional channel estimation algorithms. The complexity of SCLSTM comes from two parts, $K+1$ sparse-connected layers and $K$ LSTM blocks. The complexity of one sparse-connected layer is ${\cal O}\left( {4MN} \right)$. Thus, the total complexity of $K+1$ layers is ${\cal O}\left( {4MN+4KMN} \right)$. The complexity of the LSTM with $L$ layers is ${\cal O}\left( {{N_{LSTM}}} \right)$, where $N_{LSTM}$ is the number of parameters for the LSTM structure, which can be calculated as \cite{complexityLSTM}
	\begin{align}   \label{xu22}
	\begin{array}{l}
	{N_{LSTM}} = 4\left( {{n_{in}} \times n_c^1 + n_c^1 \times n_c^1 + n_c^1} \right)
	 + \sum\limits_{l = 2}^L {4\left( {n_c^{l - 1} \times n_c^l + n_c^l \times n_c^l + n_c^l} \right)} 
	 +  n_c^L \times {n_{out}} + {n_{out}},
	\end{array}
	\end{align}
	where ${{n_{in}}}$, ${n_{out}}$, and $n_c^l(l = 1, \cdot \cdot \cdot ,L)$ are the number of input units, output units, and memory cells, respectively. In our proposed SCLSTM, one LSTM block has two LSTM layers with $6N$ and $4N$ memory cells, respectively. Besides, a dense layer with $2N$ neurons followed by two LSTM layers. Thus, the complexity of the $K$ LSTM blocks and the SCLSTM is ${\cal O}( {360K{N^2} + 42KN} )$ and ${\cal O}( {360K{N^2} + ( {4KM+4M + 42K} )N} )$, respectively.

	\section{Simulation results}
	This section presents the simulation results to demonstrate the effectiveness of our proposed framework and the SCLSTM-based algorithm. 

	\subsection{Simulation Setup}
	We generate $\textbf{G}$ with the model in (\ref{xu4}), and ${{\textbf{h}}_k}$ with the model in (\ref{xu5}). The parameters of the two channel models are illustrated in Table \ref{parachannel}.  
	\begin{table}[tbp]
		\centering
		\caption{{Parameters of the two channel models}}
		\begin{tabular}{|c|c|c|}
			\hline
			\vspace{-0.5cm}
			&  & \\
			Parameter  &  description & value \\
			\hline
			\vspace{-0.5cm}
			&  & \\
			$M$ & The number of antennas at BS & 4 \\
			\hline
			\vspace{-0.5cm}
			&  & \\
			$K$ & The number of UEs & 4 \\
			\hline
			\vspace{-0.5cm}
			&  & \\
			$N$ &  The size of the RIS & $[20,40,60,80,100]$ \\
			\hline
			\vspace{-0.5cm}
			&  & \\
			${L_G}$ & The paths number of channel $\textbf{G}$ & 3   \\
			\hline
			\vspace{-0.5cm}
			&  & \\
			${L_k}$ & The paths number of channel ${{\textbf{h}}_k}$ & 3   \\
			\hline
			\vspace{-0.5cm}
			&  & \\
			${f_c}$ & The carrier frequency & 28 GHz   \\
			\hline
			\vspace{-0.5cm}
			&  & \\
			${v_{\max}}$ & The velocity of UEs & 3 $m/s$   \\
			\hline
		\end{tabular}
		\label{parachannel}
	\end{table}
	\begin{table}[tbp]
		\centering
		\caption{{Hyper-parameters of SCLSTM}}
		\begin{tabular}{|c|c|}
			\hline
			\vspace{-0.5cm}
			&   \\
			Parameter  & value \\
			\hline
			\vspace{-0.5cm}
			&   \\
			Training samples  & 10000 \\
			\hline
			\vspace{-0.5cm}
			&   \\
			Validation and test samples  & 1000 \\
			\hline
			\vspace{-0.5cm}
			&  \\
			Time steps & $[2,4,6,8,10]$   \\
			\hline
			\vspace{-0.5cm}
			&   \\
			Learning rate  & 0.001   \\
			\hline
			\vspace{-0.5cm}
			&   \\
			Decaying rate  & 0.00001   \\
			\hline
			\vspace{-0.5cm}
			&   \\
			Batch size  & 128   \\
			\hline
			\vspace{-0.5cm}
			&  \\
			Maximum epoches & 1000   \\
			\hline
			
		\end{tabular}
		\label{paraSCLSTM}
	\end{table}
	Moreover, the AoAs/AoDs are assumed to distribute in $[0,2\pi ]$, and the the Doppler shift ${{f_{l,k}}}$ is randomly distributed in $[0,{f_{\max }}]$, where ${f_{\max }} = \frac{{{f_c}{v_{\max }}}}{c}$ is the maximum Doppler shift, and $c$ is the velocity of light.
	
	For the different simulations, we generate training samples, validation samples, and test samples with various setups. The details are shown in Table \ref{paraSCLSTM}. 
	
	\subsection{NMSE of Cascaded Channel ${\bf{H}}$}
	For convenience, we define ${\bf{H}} = { {\bf{H}}_k  }$ and ${\bf{h}} = { {\bf{h}}_k }$. Besides, we measure the prediction accuracy of the cascaded channel in the whole simulation using the NMSE defined as 
	\begin{align}   \label{xu23}
	{\rm{NMS}}{{\rm{E}}_H} \buildrel \Delta \over = \mathbb{E}\left\{ {\frac{{\sum\limits_{k = 1}^K {{{\left\| {{{{\bf{\tilde H}}}_{k}(S + 1)} - {{\bf{H}}_{k}(S + 1)}} \right\|}^2}} }}{{\sum\limits_{k = 1}^K {{{\left\| {{{\bf{H}}_{k}(S + 1)}} \right\|}^2}} }}} \right\},
	\end{align}
	and through simple data replacement, (\ref{xu23}) is also suitable for measuring the prediction accuracy of BS-RIS channel and RIS-UE channels.
	
\begin{figure}[tbp]
		\centering {
			\begin{tabular}{ccc}
				\includegraphics[width=0.7\textwidth]{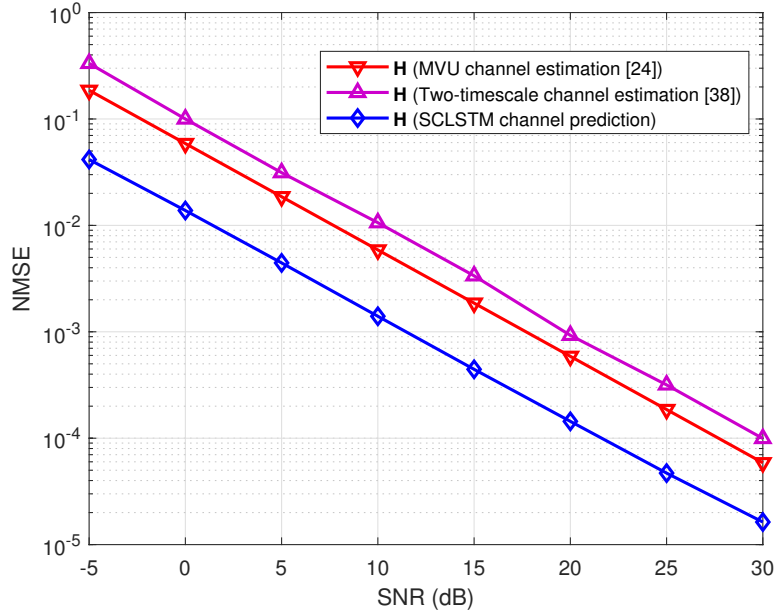}\\
			\end{tabular}
		}
		\caption{The NMSE of the cascaded channel ${\bf{H}}$ versus the uplink SNR, where $M=4$, $N=40$, $K=4$, $S=4$.}
		\vspace{-1\baselineskip}
		\label{result1}
	\end{figure}
\begin{figure}[tbp]
		\centering {
			\begin{tabular}{ccc}
				\includegraphics[width=0.7\textwidth]{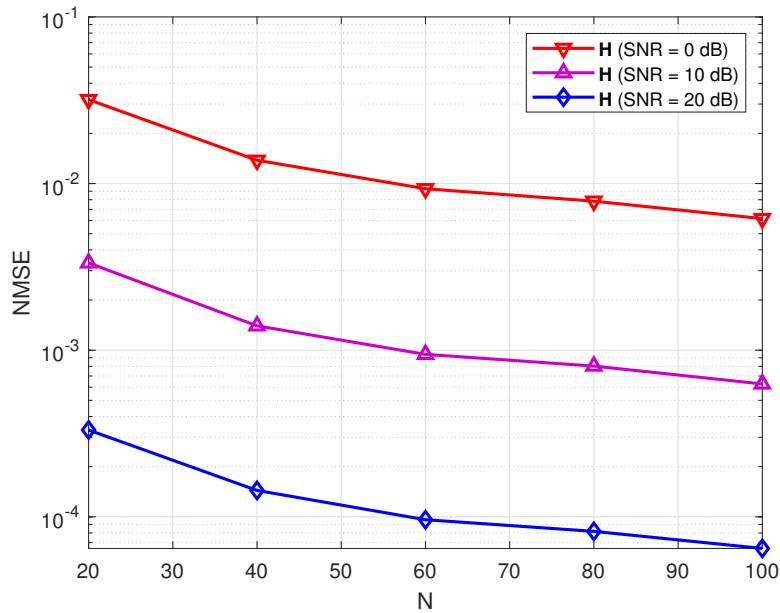}\\
			\end{tabular}
		}
		\caption{The NMSE of the cascaded channel ${\bf{H}}$ of SCLSTM versus $N$, where $M=4$, $K=4$, $S=4$.}
		\vspace{-1\baselineskip}
		\label{result2}
	\end{figure}
\begin{figure}[tbp]
		\centering {
			\begin{tabular}{ccc}
				\includegraphics[width=0.7\textwidth]{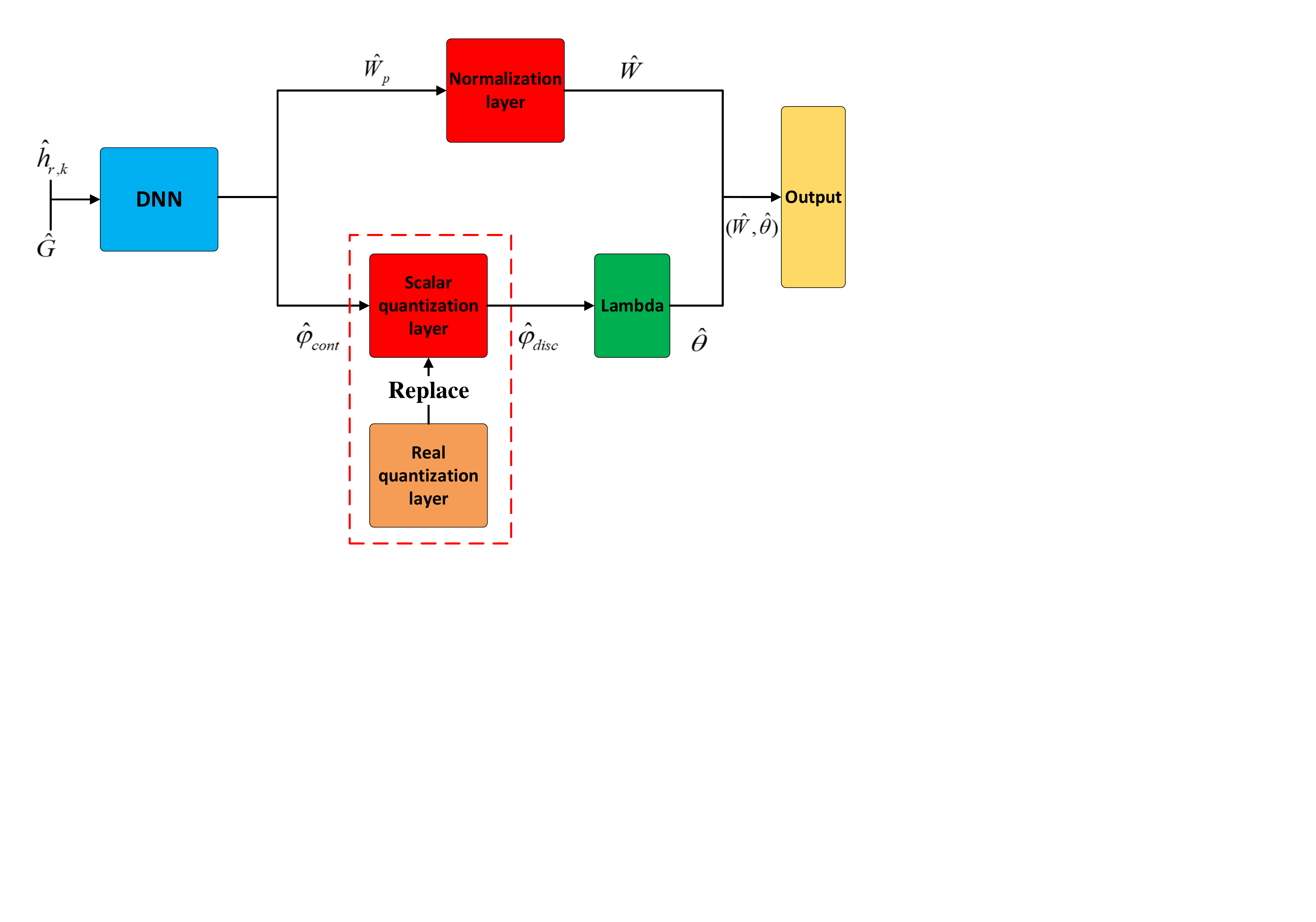}\\
			\end{tabular}
		}
		\caption{The NMSE of the cascaded channel ${\bf{H}}$ of SCLSTM versus $S$, where $M=4$, $N=40$, $K=4$.}
		\vspace{-1\baselineskip}
		\label{result3}
	\end{figure}

	The NMSE comparisons of ${\bf{H}}$ between different algorithms versus the uplink SNR are illustrated in Fig. \ref{result1} with the setup that $M=4$, $N=40$, $K=4$, and $S=4$. In Fig. \ref{result2} and Fig. \ref{result3}, we give the NMSE performance of the predicted ${\bf{H}}$ by the proposed SCLSTM-based algorithm with various $N$ and $S$. We set $M=4$, $K=4$, $S=4$ in Fig. \ref{result2}, while $M=4$, $N=40$, $K=4$ in Fig. \ref{result3}. 
	
	As shown in Fig. \ref{result1}, the NMSE of all algorithms decreases with the increase of SNR, and our proposed algorithm gets the lowest NMSE between different algorithms. The gap between the SCLSTM-based algorithm and the MVU algorithm is 6.5 dB, while it is 8 dB between the SCLSTM-based algorithm and the two-timescale algorithm. The performance of our proposed algorithm is the best one, since deep learning can eliminate the noise influence in CSI acquisition by training on a large amount of data.
	
	In Fig. \ref{result2}, the NMSE decreases with the increase of $N$. This happens because our proposed channel prediction is based on the results of the channel estimation \cite{twotime} in the second stage of our proposed framework. The reflection matrix of the RIS-UE channel estimation \cite{twotime} is designed as a DFT matrix, which causes the NMSE of the channel estimation to be inversely proportional to $N$. 
	
	Fig. \ref{result3} shows that the longer the $S$, the better the NMSE performance of the predicted ${\bf{H}}$, because the longer $S$ enables the SCLSTM to capture the temporal relationship of the time-varying channel more accurately. 
	

	\subsection{NMSE of ${\bf{G}}$ and ${\bf{h}}$}
\begin{figure}[tbp]
		\centering {
			\begin{tabular}{ccc}
				\includegraphics[width=0.7\textwidth]{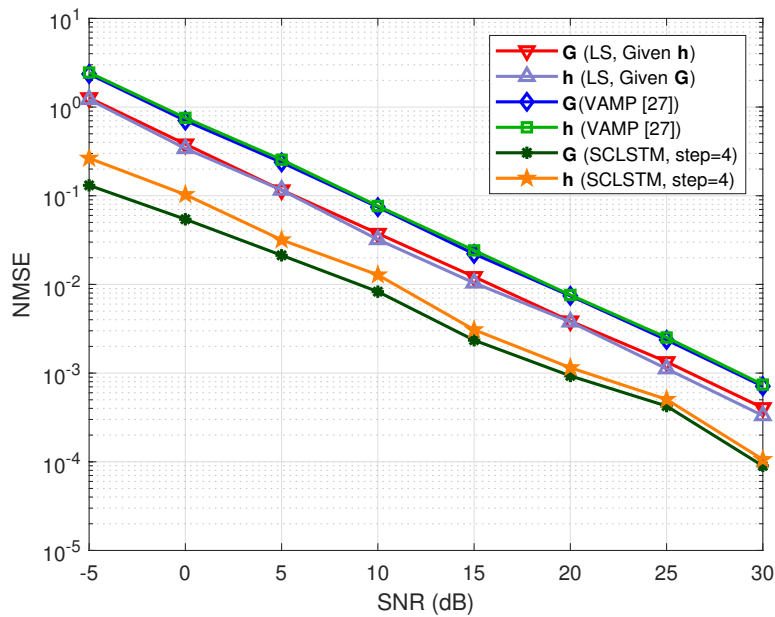}\\
			\end{tabular}
		}
		\caption{The NMSE of ${\bf{G}}$ and ${\bf{h}}$ versus the uplink SNR, where $M=4$, $N=40$, $K=4$, $S=4$.}
		\vspace{-1\baselineskip}
		\label{result4}
	\end{figure}

	Fig. \ref{result4} contains the NMSE performance comparison of ${\bf{G}}$ and ${\bf{h}}$ between the proposed SCLSTM-based channel prediction, the genie-aided LS channel estimation that estimates one unknown channel based on the another known channel, and the VAMP-based channel estimation \cite{CEhuang} with the parameter setting $M=4$, $N=40$, $K=4$, and $S=4$. As shown in Fig. \ref{result4}, there are about 5 dB and 7.5 dB NMSE performance gaps between the proposed SCLSTM-based algorithm and the genie-aided LS algorithm and the VAMP-based algorithm, respectively. For the proposed algorithm, the NMSE performance of ${\bf{h}}$ has the same result under the high SNR, but these gaps become greater under the low SNR due to the scale ambiguity correction error. 
	
	
	

	\subsection{NMSE of ${\bf{G}}$ and ${\bf{h}}$ with the Error of Estimated ${{{\bf{g}}_1}}$}
\begin{figure}[tbp]
		\centering {
			\begin{tabular}{ccc}
				\includegraphics[width=0.7\textwidth]{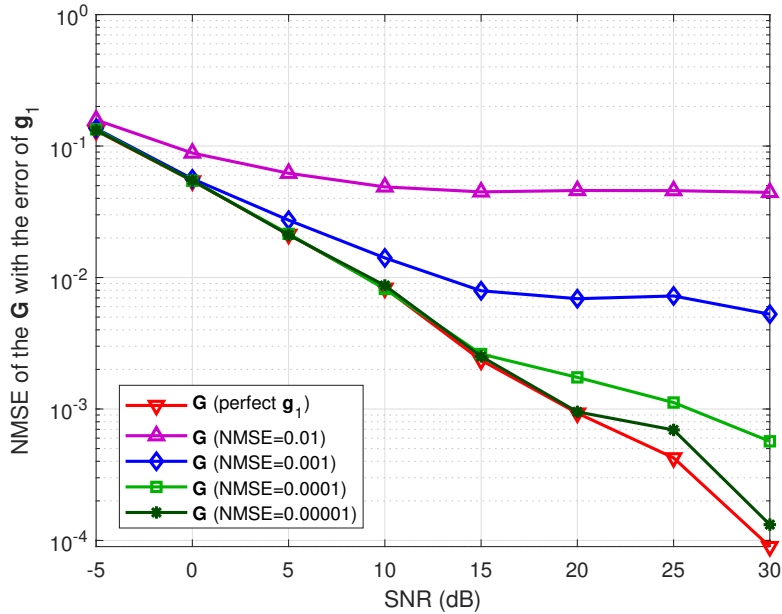}\\
			\end{tabular}
		}
		\caption{The NMSE of ${\bf{G}}$ of SCLSTM with the error of ${{\bf{g}}_1}$ versus the uplink SNR, where $M=4$, $N=40$, $K=4$, $S=4$.}
		\vspace{-1\baselineskip}
		\label{result9}
	\end{figure}

\begin{figure}[tbp]
		\centering {
			\begin{tabular}{ccc}
				\includegraphics[width=0.7\textwidth]{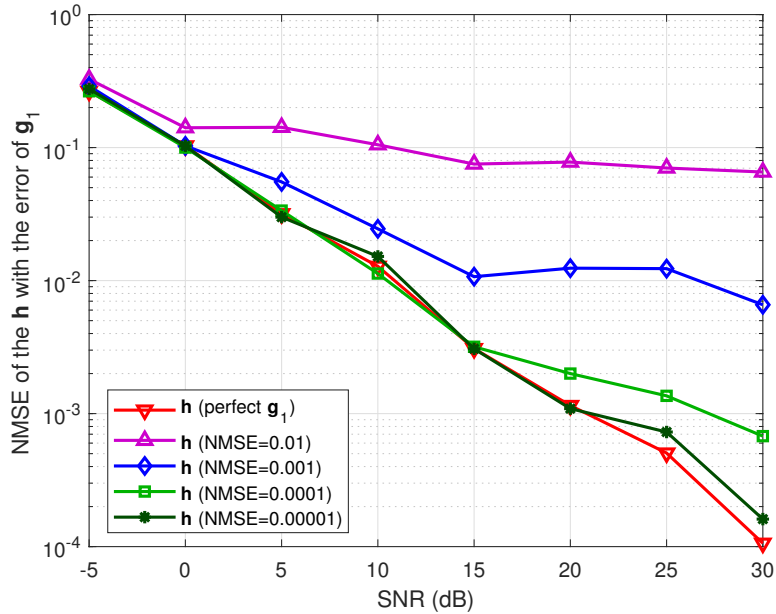}\\
			\end{tabular}
		}
		\caption{The NMSE of ${\bf{h}}$ of SCLSTM with the error of ${{\bf{g}}_1}$ versus the uplink SNR, where $M=4$, $N=40$, $K=4$, $S=4$.}
		\vspace{-1\baselineskip}
		\label{result10}
	\end{figure}

	Fig. \ref{result9} and Fig. \ref{result10} study the NMSE performance of ${\bf{G}}$ and ${\bf{h}}$ against SNR for different channel estimation errors of ${{\bf{g}}_{1}}$ under the setting $M=4$, $N=40$, $K=4$, and $S=4$, to show the impact of scaling ambiguity correction accuracy on our proposed framework. We also use NMSE to represent the channel estimation error of ${{\bf{g}}_{1}}$, marked as ${\bf{g}}_1^e$ for convenience. 
	
	Fig. \ref{result9} shows that when ${\bf{g}}_1^e = 0.01$, the NMSE of ${\bf{G}}$ has a big performance gap compared to the case of perfect ${{\bf{g}}_{1}}$, and the gap increases with the increase of SNR, e.g., when SNR = 0 dB, the gap is 2.5 dB, and when SNR = 30 dB, it increases to 30 dB. The case where ${\bf{g}}_1^e = 0.001$ is similar to that of 0.01, but the NMSE performance of ${\bf{G}}$ is improved, and when SNR $ \le $ 0 dB, the NMSE performance of ${\bf{G}}$ in the former case is the same as the perfect ${{\bf{g}}_{1}}$ case. We also observe that when SNR $ \le $ 15 dB and ${\bf{g}}_1^e = 0.0001$, the NMSE performance of ${\bf{G}}$ does not decrease compared to the case of perfect ${{\bf{g}}_{1}}$. Furthermore, while SNR $ < $ 20 dB and ${\bf{g}}_1^e = 0.00001$, the NMSE performance of ${\bf{G}}$ is the same as the case of perfect ${{\bf{g}}_{1}}$, and the gap between them is about 2 dB when SNR = 25, 30 dB. 

	We can find that the trend in Fig. \ref{result10} is the same as that in Fig. \ref{result9}. The NMSE of ${\bf{h}}$ decreases with the decrease of ${\bf{g}}_1^e$, and when the SNR is low, different ${\bf{g}}_1^e$ have minor impact on the NMSE of ${\bf{h}}$. Conversely, as the SNR increases, the smaller the ${\bf{g}}_1^e$, the smaller the NMSE of ${\bf{h}}$. When ${\bf{g}}_1^e=0.00001$, the performance similar to that of the perfect ${\bf{g}}_{1}$ can be achieved.

	\subsection{Performance of Continuous Prediction}
\begin{figure}[tbp]
		\centering {
			\begin{tabular}{ccc}
				\includegraphics[width=0.7\textwidth]{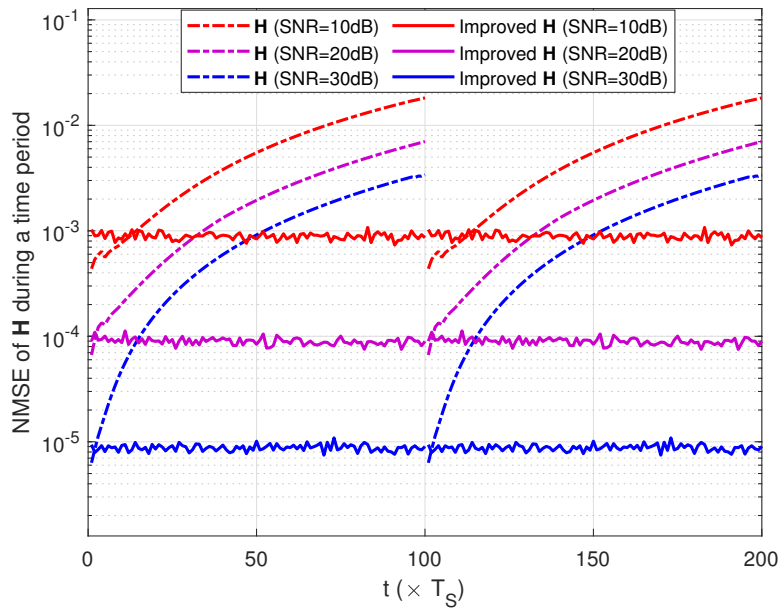}\\
			\end{tabular}
		}
		\caption{The NMSE of ${\bf{H}}$ of SCLSTM for the continues prediction versus the time, where $M=4$, $N=40$, $K=4$, $S=4$, ${T_C} = 200T_S$, and ${T_L} = 100T_S$.}
		\vspace{-1\baselineskip}
		\label{result11}
	\end{figure}

	In Fig. \ref{result11}, we plot the NMSE curve of ${\bf{H}}$ for the continuous prediction versus the time via the proposed SCLSTM-based algorithm, where $M=4$, $N=40$, $K=4$, $S=4$, ${T_C} = 200T_S$, and ${T_L} = 100T_S$. From Fig. \ref{result11}, the prediction error decreases with the increasing SNR. Moreover, when $t$ is within the same coherence time ${T_L}$, the prediction performance becomes worse over time because the prediction error at the last time accumulates to a later time. Fortunately, some improved schemes  such as the joint channel prediction and data detection can be employed to deal with this problem. Specifically, the predicted channel is first used for data detection, then known information such as the constellation diagram of the modulation mode is used to correct the detected datas, and finally, the updated datas are used as pseudo pilots to improve the channel prediction accuracy. The improved results are also shown in Fig. \ref{result11}. During a coherent time $T_L$, the improved scheme can effectively enhance the predicted channel accuracy. We also observe that the predicted channel is more accurate than the improved channel initially. This is because the improved scheme is an LS-based channel estimation with worse performance than our proposed method. In addition, we also utilize this improved scheme to the traditional pilot-based time-varying channel estimation for a fair comparison. Moreover, ${\bf{G}}$ can be obtained very accurately in the initial time step of $T_L$ and is unchanged during $T_L$. Therefore, it is also easy to get accurate ${\bf{h}}$ over time according to the predicted $G$ and the improved ${\bf{H}}$. Note that when $t$ is not in the same coherence time ${T_L}$, the large-timescale channel ${\bf{G}}$ changes, so the first and second stages of the proposed framework need to be repeated.


	\subsection{Average Downlink Sum Rate}
	To further study the impact of the CSI obtained by our proposed channel prediction algorithm and existing channel estimation algorithms on the average downlink sum rate, we use the method in \cite{guo2020weighted} to jointly optimize the precoding matrix at the BS and the reflection coefficient vector at the RIS. The average downlink sum rate is defined as
	\begin{align}   \label{xu26}
	R = \mathbb{E}\left\{ {\frac{{{T_S} - {P_a}}}{{{T_S}}}\sum\limits_{k = 1}^K {{{\log }_2}\left( {1 + {\rm{SIN}}{{\rm{R}}_{{k}}}} \right)} } \right\}, 
	\end{align}
	where ${\lambda _d} = ({T_S} - {P_a})/{T_S}$ is the data transmission coefficient, which represents the proportion of data effectively transmitted in each small-timescale $T_S$. ${{\rm{SIN}}{{\rm{R}}_{{k}}}}$ is the signal-to-interference-plus-noise ratio for the UE $k$, which is expressed as 
	\begin{align}   \label{xu27}
	{\rm{SIN}}{{\rm{R}}_{{k}}} = \frac{{{{\left| {{\bf{h}}_{k}^H{\bf{\Theta}} {\bf{G}}^H{{\bf{w}}_{k}}{x_k}} \right|}^2}}}{{\sum\limits_{n,n \ne k}^K {{{\left| {{\bf{h}}_{k}^H{\bf{{\Theta}}} {\bf{G}}^H{{\bf{w}}_{n}}{x_n}} \right|}^2} + {\sigma ^2}} }},
	\end{align}
	where ${{{\bf{w}}_{k}}}$ and ${\bf{\Theta }}$ are the precoding matrix at the BS and the reflection coefficient vector at the RIS.
	
	The data transmission coefficient and average sum rate comparisons over $T_S$ between different algorithms are shown in Fig. 13, where $M=K=S=4$, $N=40$, SNR = $20$ dB, and ${T_C} = {T_L} = 5000$.

\begin{figure}[tbp]
		\centering
		\hspace{-1\baselineskip}
		\subfigure[]{
			\label{FirstStep}
			\begin{minipage}[t]{0.5\linewidth}
				\includegraphics[width=1\textwidth]{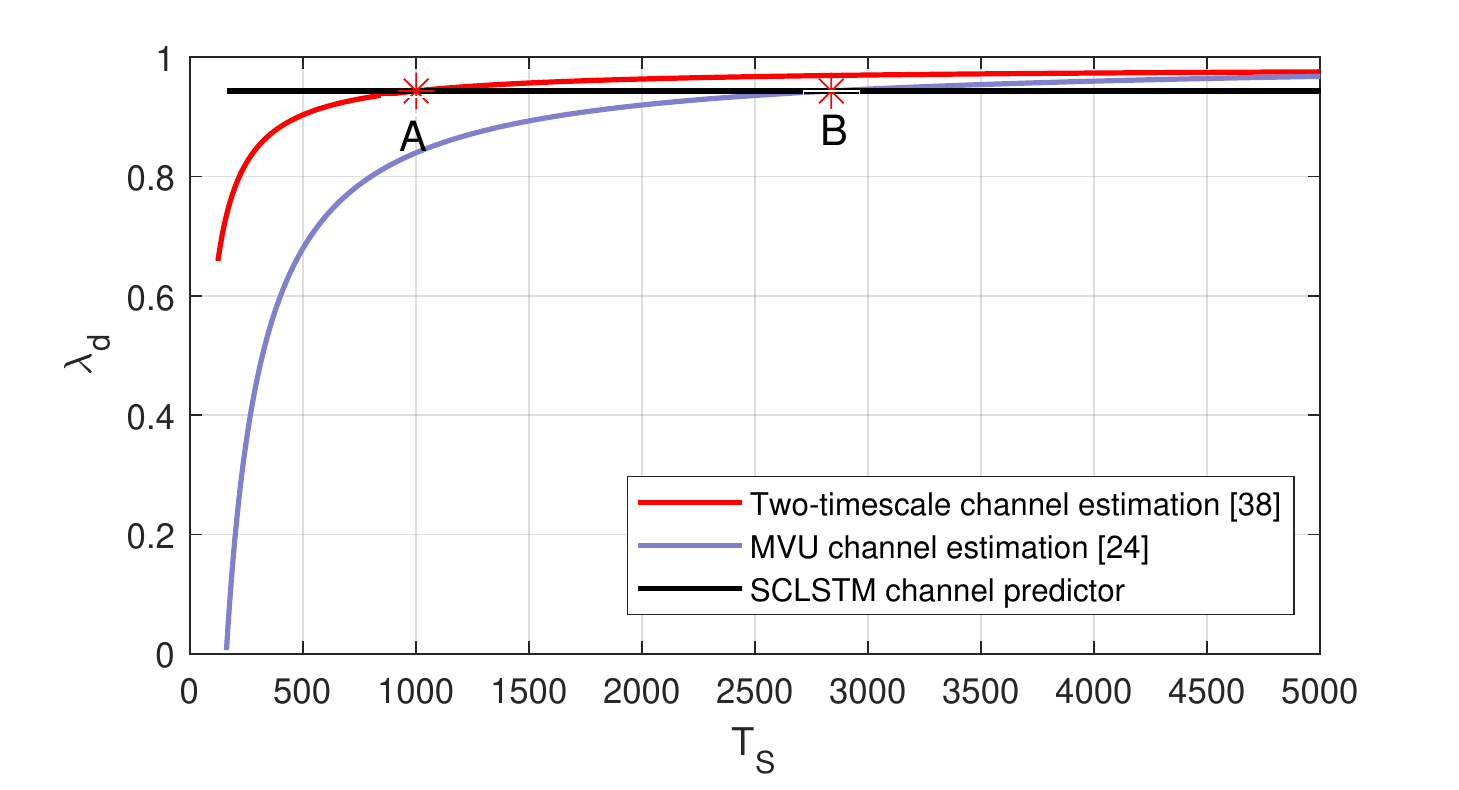}
			\end{minipage}
		}
	    \hspace{-1\baselineskip}
		\subfigure[]{
			\label{SecondStep}
			\begin{minipage}[t]{0.5\linewidth}
				\includegraphics[width=1\textwidth]{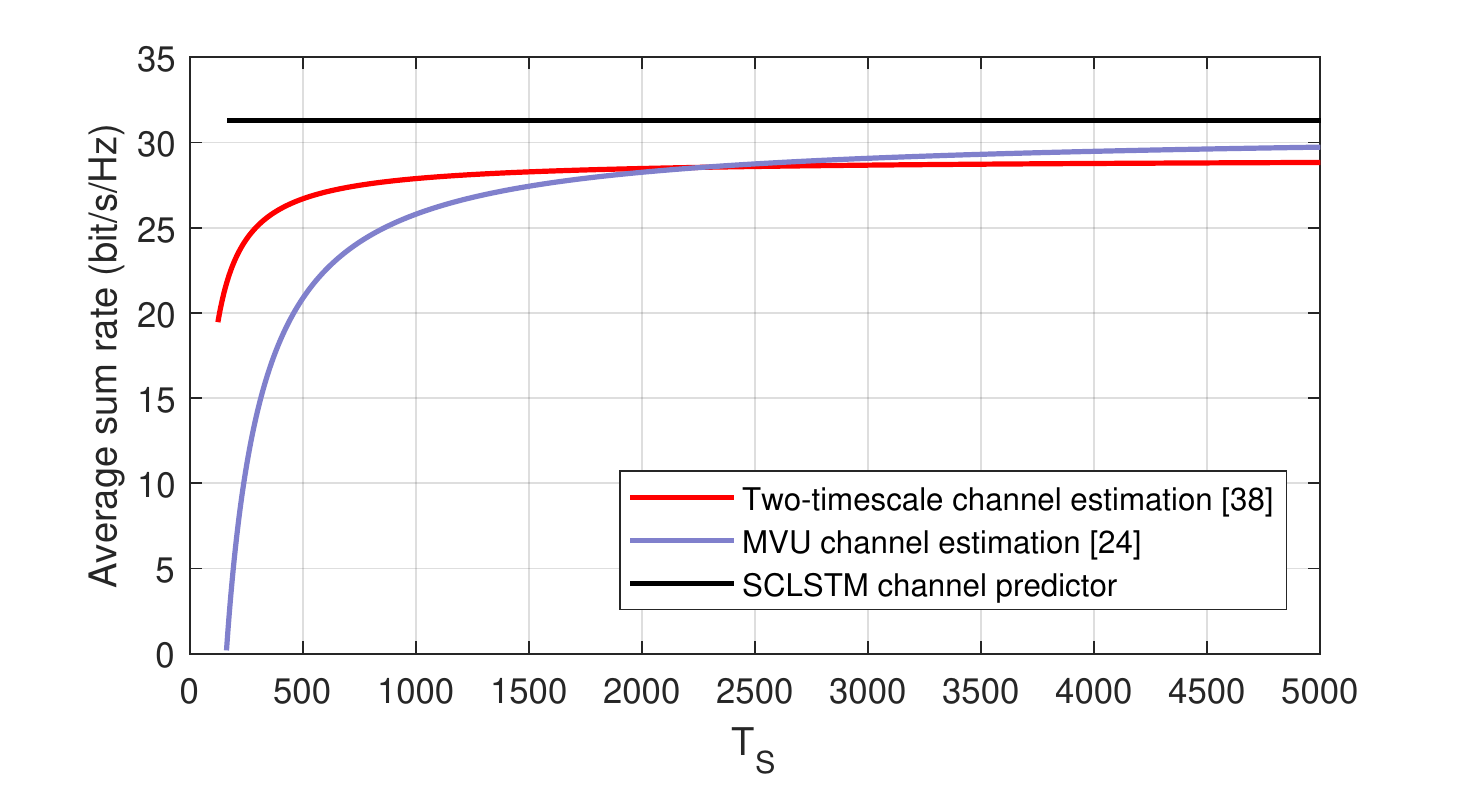}
			\end{minipage}
		}
		\caption{(a) Data transmission coefficient ${\lambda _d}$ versus $T_S$; (b) Average sum rate versus $T_S$. $M=K=S=4$, $N=40$, SNR = $20$ dB, and ${T_C} = {T_L} = 5000$.}
		\vspace{-1\baselineskip}
		\label{result12}
	\end{figure}

	
	In Fig. \ref{FirstStep}, we can find that the ${\lambda _d}$ of the SCLSTM-based algorithm is a constant, which is calculated as $1 - {P_L}/{T_L}$ with the fixed $T_L$ and $P_L$. The ${\lambda _d}$ under the MVU and two-timescale channel estimation algorithms increases with the increase of $T_S$. The lines of ${\lambda _d}$ of the SCLSTM-based algorithm and the other two algorithms have two intersection points, A $( {1000, 0.9436} )$ with $\tau = 5$ and B $(2836.88, 0.9436)$ with $\tau = 1.76$. This result is in line with our analysis of the feasible conditions of $\tau$ in proposition 1 and proposition 2. When $T_S$ is smaller than the abscissa values of A and B, respectively, the value of ${\lambda _d}$ of the SCLSTM-based algorithm is larger than that of the other algorithms. Note that the faster the channel ${\bf{h}}$ changes, the smaller the value of $T_S$ is. This shows that the proposed SCLSTM-based algorithm has higher data transmission efficiency than the existing algorithms when the channel is fast time-varying. Besides, the pilot-based channel estimation is hard to directly apply when $T_S$ is less than the period needed to send the pilot.
	
	Fig. \ref{SecondStep} shows that the average sum rate of the SCLSTM-based algorithm is always greater than that of the other two algorithms, even if $T_S=T_L$. However, ${T_S} \le {T_L}$ is always satisfied under the two-timescale assumption. Although the SCLSTM-based algorithm can have better data transmission efficiency only when $T_S$ is small, it can obtain CSI more accurately, which achieves the better average sum rate. 

	\section{Conclusion}
	In this paper, we proposed a joint time-varying channel decomposition and prediction framework based on the two-timescale property for a RIS-assisted MU-MISO wireless communication system. Different from other channel estimation algorithms, our proposed framework acquired CSI by channel prediction and effectively decomposed the BS-RIS and RIS-UE channels from the cascaded channels. Moreover, the framework employed the full-duplex scheme to estimate the channel from a specific antenna to the RIS for solving the scaling ambiguity in channel decomposition. We also proposed a novelty LSTM-based network structure, SCLSTM, and designed a corresponding algorithm for our proposed framework. Theoretical analysis demonstrated that our proposed framework had a lower pilot overhead and computational complexity by comparing it with the existing channel estimation algorithms. Simulation results verified that the proposed algorithm performed better than other traditional channel estimation algorithms and was robust to the estimation error of all channels in the first and second stages used to correct the scaling ambiguity and predict, respectively.
	
	\begin{appendices}
		\section{Proof of proposition 1}
		Referring to Table \ref{Pilot}, when $P_a$ of our proposed framework is lower than that of the MVU algorithm and the PARAFAC-VAMP algorithm, the following formulas need to be satisfied.
		\begin{align}   \label{proof11}
		\frac{{3N + 2 + KS\lceil {\frac{N}{M}} \rceil }}{\tau } \le NK.
		\end{align}
		\begin{align}   \label{proof12}
		\frac{{3N + 2 + KS\lceil {\frac{N}{M}} \rceil }}{\tau } \le KMP.
		\end{align}
		
		(\ref{proof12}) can be relaxed as (\ref{proof11}) because $MP \ge N$ is needed for efficient channel estimation from the simulation results of \cite{CEhuang}. Then, (\ref{proof11}) becomes 
		\begin{align}   \label{proof13}
		\tau \ge \frac{{3N + 2 + KS\lceil {\frac{N}{M}} \rceil }}{{NK}}.
		\end{align}
		Since $\lceil {\frac{N}{M}} \rceil  < \frac{N}{M} + 1$ holds, (\ref{proof13}) can be rewritten as 
		\begin{align}   \label{proof14}
		\tau  &> \frac{{3N + 2 + KS( {\frac{N}{M} + 1} )}}{{NK}}\notag\\
		&= \frac{3}{K} + \frac{S}{M} + \frac{S}{N} + \frac{2}{{NK}}.
		\end{align} 
		The proof is completed.
		
	\end{appendices}

	\ifCLASSOPTIONcaptionsoff
	\newpage
	\fi
	
	\bibliographystyle{IEEEtran}
	\bibliography{IEEEref}

\begin{thebibliography}{10}
\providecommand{\url}[1]{#1}
\csname url@samestyle\endcsname
\providecommand{\newblock}{\relax}
\providecommand{\bibinfo}[2]{#2}
\providecommand{\BIBentrySTDinterwordspacing}{\spaceskip=0pt\relax}
\providecommand{\BIBentryALTinterwordstretchfactor}{4}
\providecommand{\BIBentryALTinterwordspacing}{\spaceskip=\fontdimen2\font plus
\BIBentryALTinterwordstretchfactor\fontdimen3\font minus
  \fontdimen4\font\relax}
\providecommand{\BIBforeignlanguage}[2]{{%
\expandafter\ifx\csname l@#1\endcsname\relax
\typeout{** WARNING: IEEEtran.bst: No hyphenation pattern has been}%
\typeout{** loaded for the language `#1'. Using the pattern for}%
\typeout{** the default language instead.}%
\else
\language=\csname l@#1\endcsname
\fi
#2}}
\providecommand{\BIBdecl}{\relax}
\BIBdecl

\bibitem{B5G}
K.~Samdanis and T.~Taleb, ``The road beyond 5g: A vision and insight of the key
  technologies,'' \emph{IEEE Network}, vol.~34, no.~2, pp. 135--141, Feb. 2020.

\bibitem{6G2}
G.~Gui, M.~Liu, F.~Tang, N.~Kato, and F.~Adachi, ``6g: Opening new horizons for
  integration of comfort, security, and intelligence,'' \emph{IEEE Wireless
  Communications}, vol.~27, no.~5, pp. 126--132, Oct. 2020.

\bibitem{larsson2014massive}
E.~G. Larsson, O.~Edfors, F.~Tufvesson, and T.~L. Marzetta, ``Massive {MIMO}
  for next generation wireless systems,'' \emph{IEEE Communications Magazine},
  vol.~52, no.~2, pp. 186--195, Feb. 2014.

\bibitem{liangrelay}
L.~Zhang, J.~Liu, M.~Xiao, G.~Wu, Y.-C. Liang, and S.~Li, ``Performance
  analysis and optimization in downlink {NOMA} systems with cooperative
  full-duplex relaying,'' \emph{IEEE Journal on Selected Areas in
  Communications}, vol.~35, no.~10, pp. 2398--2412, Oct. 2017.

\bibitem{neikato2}
H.~Nishiyama, M.~Ito, and N.~Kato, ``Relay-by-smartphone: realizing multihop
  device-to-device communications,'' \emph{IEEE Communications Magazine},
  vol.~52, no.~4, pp. 56--65, Apr. 2014.

\bibitem{huangTWC}
C.~Huang, A.~Zappone, G.~C. Alexandropoulos, M.~Debbah, and C.~Yuen,
  ``Reconfigurable intelligent surfaces for energy efficiency in wireless
  communication,'' \emph{IEEE Transactions on Wireless Communications},
  vol.~18, no.~8, pp. 4157--4170, Aug. 2019.

\bibitem{liangyingchang1}
S.~Gong, X.~Lu, D.~T. Hoang, D.~Niyato, L.~Shu, D.~I. Kim, and Y.-C. Liang,
  ``Toward smart wireless communications via intelligent reflecting surfaces: A
  contemporary survey,'' \emph{IEEE Communications Surveys Tutorials}, vol.~22,
  no.~4, pp. 2283--2314, Jun. 2020.

\bibitem{wu2019beamforming}
Q.~Wu and R.~Zhang, ``Beamforming optimization for wireless network aided by
  intelligent reflecting surface with discrete phase shifts,'' \emph{IEEE
  Transactions on Communications}, vol.~68, no.~3, pp. 1838--1851, Mar. 2020.

\bibitem{Hoang1}
B.~Lyu, P.~Ramezani, D.~T. Hoang, S.~Gong, Z.~Yang, and A.~Jamalipour,
  ``Optimized energy and information relaying in self-sustainable
  {IRS}-empowered {WPCN},'' \emph{IEEE Transactions on Communications},
  vol.~69, no.~1, pp. 619--633, Jan. 2021.

\bibitem{WCT}
Y.~Xu, Z.~Gao, Z.~Wang, C.~Huang, Z.~Yang, and C.~Yuen, ``{RIS}-enhanced wpcns:
  Joint radio resource allocation and passive beamforming optimization,''
  \emph{IEEE Transactions on Vehicular Technology}, vol.~70, no.~8, pp.
  7980--7991, Aug. 2021.

\bibitem{huang2020holographic}
C.~Huang, S.~Hu, G.~C. Alexandropoulos, A.~Zappone, C.~Yuen, R.~Zhang,
  M.~Di~Renzo, and M.~Debbah, ``Holographic {{MIMO}} surfaces for 6g wireless
  networks: Opportunities, challenges, and trends,'' \emph{IEEE Wireless
  Communications}, vol.~27, no.~5, pp. 118--125, Oct. 2020.

\bibitem{outage}
Z.~Gao, Y.~Xu, Q.~Wang, Q.~Wu, and D.~Li, ``Outage-constrained energy
  efficiency maximization for {RIS}-assisted {WPCN}s,'' \emph{IEEE
  Communications Letters}, early access, 2021,
  doi:{10.1109/LCOMM.2021.3101657}.

\bibitem{CST}
Y.~Xu, G.~Gui, H.~Gacanin, and F.~Adachi, ``A survey on resource allocation for
  5g heterogeneous networks: Current research, future trends, and challenges,''
  \emph{IEEE Communications Surveys Tutorials}, vol.~23, no.~2, pp. 668--695,
  Feb. 2021.

\bibitem{zhiqin}
Y.~Guo, Z.~Qin, Y.~Liu, and N.~Al-Dhahir, ``Intelligent reflecting surface
  aided multiple access over fading channels,'' \emph{IEEE Transactions on
  Communications}, vol.~69, no.~3, pp. 2015--2027, Mar. 2021.

\bibitem{gaofeifei1}
S.~Zhang, S.~Zhang, F.~Gao, J.~Ma, and O.~A. Dobre, ``Deep learning optimized
  sparse antenna activation for reconfigurable intelligent surface assisted
  communication,'' \emph{IEEE Transactions on Communications}, 2021,
  doi:{10.1109/TCOMM.2021.3097726}.

\bibitem{energyefficient}
A.~Zappone, M.~Di~Renzo, F.~Shams, X.~Qian, and M.~Debbah, ``Overhead-aware
  design of reconfigurable intelligent surfaces in smart radio environments,''
  \emph{IEEE Transactions on Wireless Communications}, vol.~20, no.~1, pp.
  126--141, Jan. 2021.

\bibitem{wujointcontinuous}
Q.~Wu and R.~Zhang, ``Intelligent reflecting surface enhanced wireless network
  via joint active and passive beamforming,'' \emph{IEEE Transactions on
  Wireless Communications}, vol.~18, no.~11, pp. 5394--5409, Nov. 2019.

\bibitem{guo2020weighted}
H.~Guo, Y.-C. Liang, J.~Chen, and E.~G. Larsson, ``Weighted sum-rate
  maximization for reconfigurable intelligent surface aided wireless
  networks,'' \emph{IEEE Transactions on Wireless Communications}, vol.~19,
  no.~5, pp. 3064--3076, Feb. 2020.

\bibitem{10MISOCEBO}
Q.~{Nadeem}, H.~{Alwazani}, A.~{Kammoun}, A.~{Chaaban}, M.~{Debbah}, and
  M.~{Alouini}, ``Intelligent reflecting surface-assisted multi-user {MISO}
  communication: Channel estimation and beamforming design,'' \emph{IEEE Open
  Journal of the Communications Society}, vol.~1, pp. 661--680, May 2020.

\bibitem{an1}
J.~An and L.~Gan, ``The low-complexity design and optimal training overhead for
  {IRS}-assisted {MISO} systems,'' \emph{IEEE Wireless Communications Letters},
  vol.~10, no.~8, pp. 1820--1824, Aug. 2021.

\bibitem{an2}
J.~An, C.~Xu, L.~Wang, Y.~Liu, L.~Gan, and L.~Hanzo, ``Joint training of the
  superimposed direct and reflected links in reconfigurable intelligent surface
  assisted multiuser communications,'' \emph{arXiv preprint arXiv:2105.14484},
  2021.

\bibitem{13xu}
C.~{Huang}, G.~C. {Alexandropoulos}, A.~{Zappone}, M.~{Debbah}, and C.~{Yuen},
  ``Energy efficient multi-user {MISO} communication using low resolution large
  intelligent surfaces,'' in \emph{Proc. IEEE Globecom Workshops (GC Wkshps)},
  Abu Dhabi, United Arab Emirates, Dec. 2018, pp. 1--6.

\bibitem{14xu}
H.~{Guo}, Y.~{Liang}, J.~{Chen}, and E.~G. {Larsson}, ``Weighted sum-rate
  maximization for intelligent reflecting surface enhanced wireless networks,''
  in \emph{Proc. IEEE Global Communications Conference (GLOBECOM)}, Waikoloa,
  HI, USA, Feb. 2019, pp. 1--6.

\bibitem{15xu}
Q.~{Wu} and R.~{Zhang}, ``Beamforming optimization for wireless network aided
  by intelligent reflecting surface with discrete phase shifts,'' \emph{IEEE
  Transactions on Communications}, vol.~68, no.~3, pp. 1838--1851, Dec. 2020.

\bibitem{di2020hybrid}
B.~Di, H.~Zhang, L.~Song, Y.~Li, Z.~Han, and H.~V. Poor, ``Hybrid beamforming
  for reconfigurable intelligent surface based multi-user communications:
  Achievable rates with limited discrete phase shifts,'' \emph{IEEE Journal on
  Selected Areas in Communications}, vol.~38, no.~8, pp. 1809--1822, Jun. 2020.

\bibitem{NOMA}
T.~Hou, Y.~Liu, Z.~Song, X.~Sun, Y.~Chen, and L.~Hanzo, ``Reconfigurable
  intelligent surface aided {NOMA} networks,'' \emph{IEEE Journal on Selected
  Areas in Communications}, vol.~38, no.~11, pp. 2575--2588, Nov. 2020.

\bibitem{mmwavethz}
Z.~Wan, Z.~Gao, F.~Gao, M.~D. Renzo, and M.-S. Alouini, ``Terahertz massive
  {MIMO} with holographic reconfigurable intelligent surfaces,'' \emph{IEEE
  Transactions on Communications}, vol.~69, no.~7, pp. 4732--4750, Jul. 2021.

\bibitem{cascaded1}
D.~Mishra and H.~Johansson, ``Channel estimation and low-complexity beamforming
  design for passive intelligent surface assisted {MISO} wireless energy
  transfer,'' in \emph{Proc. IEEE International Conference on Acoustics, Speech
  and Signal Processing (ICASSP)}, Brighton, UK, May 2019, pp. 4659--4663.

\bibitem{cascaded2}
Z.-Q. He and X.~Yuan, ``Cascaded channel estimation for large intelligent
  metasurface assisted massive {MIMO},'' \emph{IEEE Wireless Communications
  Letters}, vol.~9, no.~2, pp. 210--214, Feb. 2020.

\bibitem{MVU}
T.~L. {Jensen} and E.~{De Carvalho}, ``An optimal channel estimation scheme for
  intelligent reflecting surfaces based on a minimum variance unbiased
  estimator,'' in \emph{Proc. IEEE International Conference on Acoustics,
  Speech and Signal Processing (ICASSP)}, Barcelona, Spain, May 2020, pp.
  5000--5004.

\bibitem{BSRIS}
Z.~Wang, L.~Liu, and S.~Cui, ``Channel estimation for intelligent reflecting
  surface assisted multiuser communications: Framework, algorithms, and
  analysis,'' \emph{IEEE Transactions on Wireless Communications}, vol.~19,
  no.~10, pp. 6607--6620, Oct. 2020.

\bibitem{CStradition}
K.~Ardah, S.~Gherekhloo, A.~L.~F. de~Almeida, and M.~Haardt, ``Trice: A channel
  estimation framework for {RIS}-aided millimeter-wave {MIMO} systems,''
  \emph{IEEE Signal Processing Letters}, vol.~28, pp. 513--517, Feb. 2021.

\bibitem{CEhuang}
L.~Wei, C.~Huang, G.~C. Alexandropoulos, C.~Yuen, Z.~Zhang, and M.~Debbah,
  ``Channel estimation for {RIS}-empowered multi-user {MISO} wireless
  communications,'' \emph{IEEE Transactions on Communications}, vol.~69, no.~6,
  pp. 4144--4157, Mar. 2021.

\bibitem{CEDL1}
C.~{Huang}, G.~C. {Alexandropoulos}, C.~{Yuen}, and M.~{Debbah}, ``Indoor
  signal focusing with deep learning designed reconfigurable intelligent
  surfaces,'' in \emph{Proc. IEEE International Workshop on Signal Processing
  Advances in Wireless Communications (SPAWC)}, Cannes, France, Jul 2019, pp.
  1--5.

\bibitem{CEDL2}
G.~C. Alexandropoulos and E.~Vlachos, ``A hardware architecture for
  reconfigurable intelligent surfaces with minimal active elements for explicit
  channel estimation,'' in \emph{Proc. IEEE International Conference on
  Acoustics, Speech and Signal Processing (ICASSP)}, Barcelona, Spain, May
  2020, pp. 9175--9179.

\bibitem{CEDL3}
A.~Taha, M.~Alrabeiah, and A.~Alkhateeb, ``Enabling large intelligent surfaces
  with compressive sensing and deep learning,'' \emph{IEEE Access}, vol.~9, pp.
  44\,304--44\,321, Mar. 2021.

\bibitem{CEDL4}
A.~M. Elbir, A.~Papazafeiropoulos, P.~Kourtessis, and S.~Chatzinotas, ``Deep
  channel learning for large intelligent surfaces aided mm-wave massive {MIMO}
  systems,'' \emph{IEEE Wireless Communications Letters}, vol.~9, no.~9, pp.
  1447--1451, Sep. 2020.

\bibitem{CP1}
A.~Duel-Hallen, ``Fading channel prediction for mobile radio adaptive
  transmission systems,'' \emph{Proceedings of the IEEE}, vol.~95, no.~12, pp.
  2299--2313, Dec. 2007.

\bibitem{CP2}
K.~Baddour and N.~Beaulieu, ``Autoregressive modeling for fading channel
  simulation,'' \emph{IEEE Transactions on Wireless Communications}, vol.~4,
  no.~4, pp. 1650--1662, Jul. 2005.

\bibitem{CP3}
L.~Fan, Q.~Wang, Y.~Huang, and L.~Yang, ``Performance analysis of
  low-complexity channel prediction for uplink massive {MIMO},'' \emph{IET
  Communications}, vol.~10, no.~14, pp. 1744--1751, Sep. 2016.

\bibitem{CP4}
C.~Min, N.~Chang, J.~Cha, and J.~Kang, ``{MIMO}-{OFDM} downlink channel
  prediction for {IEEE}802.16e systems using {K}alman filter,'' in \emph{Proc.
  IEEE Wireless Communications and Networking Conference (WCNC)}, Hong Kong,
  China, Jun. 2007, pp. 942--946.

\bibitem{wangSVmodel}
P.~Wang, J.~Fang, H.~Duan, and H.~Li, ``Compressed channel estimation for
  intelligent reflecting surface-assisted millimeter wave systems,'' \emph{IEEE
  Signal Processing Letters}, vol.~27, pp. 905--909, May 2020.

\bibitem{TSVmodel}
R.~W. Heath, N.~González-Prelcic, S.~Rangan, W.~Roh, and A.~M. Sayeed, ``An
  overview of signal processing techniques for millimeter wave {MIMO}
  systems,'' \emph{IEEE Journal of Selected Topics in Signal Processing},
  vol.~10, no.~3, pp. 436--453, Feb. 2016.

\bibitem{twotime}
C.~Hu, L.~Dai, S.~Han, and X.~Wang, ``Two-timescale channel estimation for
  reconfigurable intelligent surface aided wireless communications,''
  \emph{IEEE Transactions on Communications}, early access, 2021, doi:
  {10.1109/TCOMM.2021.3072729}.

\bibitem{huyixing}
S.~Atapattu, R.~Fan, P.~Dharmawansa, G.~Wang, J.~Evans, and T.~A. Tsiftsis,
  ``Reconfigurable intelligent surface assisted two–way communications:
  Performance analysis and optimization,'' \emph{IEEE Transactions on
  Communications}, vol.~68, no.~10, pp. 6552--6567, Jul. 2020.

\bibitem{LSTM1}
S.~Hochreiter and J.~Schmidhuber, ``Long short-term memory,'' \emph{Neural
  computation}, vol.~9, no.~8, pp. 1735--1780, Nov. 1997.

\bibitem{chidu2}
Y.~Rong, M.~R.~A. Khandaker, and Y.~Xiang, ``Channel estimation of dual-hop
  {MIMO} relay system via parallel factor analysis,'' \emph{IEEE Transactions
  on Wireless Communications}, vol.~11, no.~6, pp. 2224--2233, Apr. 2012.

\bibitem{complexityLSTM}
W.~Jiang and H.~D. Schotten, ``Deep learning for fading channel prediction,''
  \emph{IEEE Open Journal of the Communications Society}, vol.~1, pp. 320--332,
  Mar. 2020.

\end{thebibliography}

\end{document}